\documentclass[11pt,a4paper]{article}
\usepackage[T1]{fontenc}
\usepackage[utf8]{inputenc}
\usepackage{lmodern}
\usepackage{microtype}
\usepackage[margin=25mm]{geometry}
\usepackage{amsmath,amssymb}
\usepackage{booktabs}
\usepackage{graphicx}
\usepackage{placeins}
\usepackage[round,authoryear]{natbib}
\usepackage[hidelinks]{hyperref}
\hypersetup{pdftitle={Bitcoin UTXO Values Are Heavy-Tailed but Not Power-Law Distributed: A Clauset-Shalizi-Newman Test across Eight Snapshots}, pdfauthor={Carlos Baquero and Raquel Menezes}, pdfsubject={Preprint, September 2026}, pdfkeywords={Bitcoin, UTXO, power law, heavy tails, Clauset-Shalizi-Newman, goodness of fit, bootstrap}}
\title{Bitcoin UTXO Values Are Heavy-Tailed\\but Not Power-Law Distributed\\
\large A Clauset--Shalizi--Newman Test across Eight Snapshots}
\author{Carlos Baquero\thanks{Faculty of Engineering, University of Porto and INESC TEC, Portugal. \texttt{cbm@fe.up.pt}}
  \and Raquel Menezes\thanks{Centro de Matem\'{a}tica, Universidade do Minho, Braga, Portugal. \texttt{rmenezes@math.uminho.pt}}}
\date{Preprint, 28 September 2026}

\begin{document}
\maketitle

\begin{abstract}
Power laws are often claimed for Bitcoin quantities from the slope of a log--log plot. We test one such claim with a formal protocol: whether the upper tail of the values of live unspent transaction outputs (UTXOs) follows a power law. We reconstruct the exact UTXO set at eight block heights from 2012 to 2025, validate it against an independent Bitcoin Core snapshot, and apply the Clauset--Shalizi--Newman protocol with three adaptations the data require: exact discrete support, several near-tied cutoffs retained and reselected in every bootstrap replicate, and resampling by transaction, because the outputs of one transaction are not independent. The protocol was fixed before inference and every later change is recorded. The pure power law is rejected at every height, with observed Kolmogorov--Smirnov distances five to forty times those the fitted law produces, and where the exponent is below 2 its extrapolation predicts, in expectation, several outputs each larger than the total coin supply. A lognormal, a stretched exponential and a power law with an exponential cutoff near a thousand bitcoin fit better in sample at every height, as families that contain the power law must, by margins that exceed those seen on simulated power-law tails at six of the eight heights, and they predict held-out transactions better at nearly every height. They rank differently from height to height, and the lognormal and the cutoff power law differ clearly at only two. The tail's shape and the share of supply it holds drifted slowly while the number of outputs grew seventy-fold. The rejection survives a null that reproduces the dependence measured within transactions, which halves the one exceptional effect size in the panel and changes no verdict. Bitcoin UTXO values are heavy-tailed but not power-law distributed.
\end{abstract}

\section{Introduction}
\label{sec:intro}

Claims that some quantity in Bitcoin follows a power law have accompanied the system for most of its history. The best known concerns price: on logarithmic axes, the dollar price against time since the genesis block lies close to a straight line over six orders of magnitude, an observation popularised as a ``power-law corridor'' \citep{Burger2019} and given a mechanistic derivation by \citet{SantostasiPerrenod2026}. Others concern the ledger itself. \citet{Kondor2014} reported power-law degree distributions for the transaction network and found a stretched exponential preferable to a power law for address balances. Later network studies described scale-free degrees in some graph constructions \citep{Aspembitova2019} and rejected them in others \citep{Liang2018}; see also \citet{LischkeFabian2016} and \citet{Maesa2018} on the structure of the early network and its users graph. \citet{SornetteZhang2025} reported power-law holding times, \citet{Li2019} a double power law with exponential decay for the sizes of on-chain transfers, and \citet{Park2026} geometric occupancy statistics for round UTXO denominations. Balances of address clusters, meanwhile, have been described as lognormal \citep{Zhang2025}. The literature agrees that these distributions are heavy-tailed and disagrees on their form.

\begin{figure}[t]
\centering
\includegraphics[width=\textwidth]{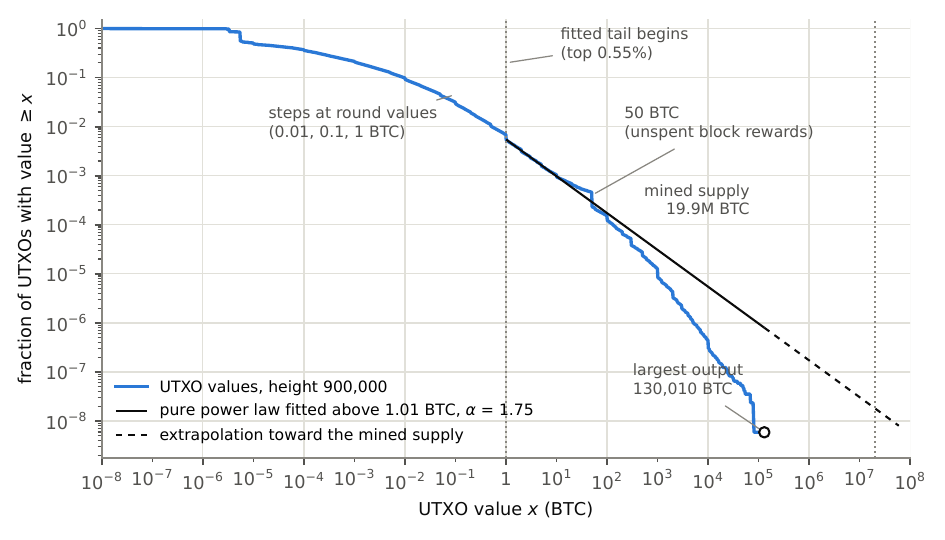}
\caption{Complementary cumulative distribution of the value of every unspent transaction output at block height 900{,}000 (June 2025; 169.8~million outputs), on logarithmic axes. The solid black line is the pure power law fitted above the population cutoff of 1.01~BTC as described in Section~\ref{sec:methods} ($\alpha = 1.75$, the top 0.55\% of outputs); the dashed line extrapolates it. The data fall below the line by two orders of magnitude before the largest existing output, 130{,}010~BTC, while the extrapolation still carries mass beyond the 19.9~million BTC mined by that height. Vertical steps mark round denominations; the step at 50~BTC is unspent block rewards.}
\label{fig:ccdf900k}
\end{figure}

Part of the disagreement is methodological. A straight line on log--log axes is weak evidence. Lognormal, stretched-exponential and truncated power-law distributions produce convincing straight segments over several orders of magnitude \citep{Mitzenmacher2004,Newman2005}, and most published power laws lack statistical support \citep{StumpfPorter2012}. \citet{Clauset2009} made the problem precise and gave a protocol: estimate the lower cutoff by minimising the Kolmogorov--Smirnov distance, estimate the exponent by maximum likelihood, obtain a goodness-of-fit $p$-value from a semiparametric bootstrap that re-estimates both, and compare the power law with alternatives by likelihood ratio. Applied at scale to network degree sequences, the protocol found pure power laws to be rare \citep{BroidoClauset2019}. No Bitcoin power-law claim we are aware of has been put through it with the bootstrap step and with the dependence in the data accounted for. Figure~\ref{fig:ccdf900k} shows the pattern this paper tests: a stretch that looks straight on logarithmic axes, and a tail that bends away from it.

Which Bitcoin quantity to test is not obvious. The protocol tests a distribution: many draws of one random variable. Most quantities called power laws in Bitcoin are time series, price, hash rate or the count of unspent outputs against time. These are ordered, trending and dependent, and the protocol does not transfer to them; the methodology they require is future work. Among cross-sectional quantities, address balances aggregate outputs by locking script and equate an address with an owner, which address reuse and custodial pooling contradict; fees and transaction sizes follow policy regimes; degree distributions depend on how the graph is built. The value of an unspent transaction output (UTXO) has none of these problems.\footnote{New bitcoin enters the ledger only through the coinbase transaction that opens each block, which pays the current block subsidy together with the fees of the transactions the block confirms; the outputs it creates carry a coinbase flag and cannot be spent until a hundred further blocks have been mined. Every other output is created by a transaction that spends outputs already in the ledger. An output is locked by a script, in the common forms to a public key or its hash, and spending it requires a signature valid under the corresponding private key, so an output is controlled by whoever holds that key.} It is the ledger's own accounting object, an exact integer number of satoshis;\footnote{One bitcoin (BTC) is $10^{8}$ satoshis, the smallest unit the protocol records. Every value in this paper is an integer number of satoshis; we quote values in BTC where that is easier to read.} the full population is observable at any block height; and anyone with a node can reproduce a snapshot bit for bit. A UTXO is not a person or a holding, and we make no claim about ownership.

We therefore ask one question: is the upper tail of the distribution of live UTXO values consistent with a power law, and is the answer stable across the blockchain's history? No prior study has fitted and tested the distribution of live UTXO values, apart from a preliminary analysis by the authors on balances binned into nine logarithmic buckets \citep{BaqueroMenezes2026}, which the present study supersedes. \citet{Jourdan2018} plotted the values of transaction outputs in an entity-labelled subset of the 2016 to 2018 ledger and described the plot as qualitatively power-law between $10^{-3}$ and $10^{3}$~BTC, without a fit or a test; the nearest fitted results are a power-law tail for address balances that its authors then found a stretched exponential described better \citep{Kondor2014}, and a double power law with exponential decay for on-chain transfer sizes \citep{Li2019}. We reconstruct the exact UTXO set at eight block heights from 2012 to 2025, validate the reconstruction against an independent Bitcoin Core snapshot, and apply the protocol with three adaptations the data require: exact discrete values; several near-tied candidate cutoffs, all retained and re-selected in every bootstrap replicate; and resampling by transaction, because outputs of one transaction are not independent draws. The analysis protocol was fixed before inference and every later change is recorded as a dated amendment (Appendix~\ref{app:protocol}).

The answer is the same at every height. A pure power law is rejected at every height, and at the seven heights where the fitted exponent is below 2 its extrapolation predicts, in expectation, several outputs each larger than the total coin supply. The tail is heavy, with a local exponent that rises from about 1.5 above 0.1~BTC to about 2.2 above 100~BTC, and it is better described by a power law with an exponential cutoff near a thousand bitcoin, by a lognormal and by a stretched exponential; which of the three fits best depends on the height, and at six of the eight the data give no clear evidence of a difference between the first two. Section~\ref{sec:data} describes the data, Section~\ref{sec:methods} the protocol and its adaptations, Section~\ref{sec:results} the results, Section~\ref{sec:discussion} their interpretation and limits, Section~\ref{sec:related} how this differs from earlier work, and Section~\ref{sec:conclusion} concludes.

\section{Data}
\label{sec:data}

\paragraph{Observation unit.}
Bitcoin keeps no balances. Its ledger is a set of unspent transaction outputs (UTXOs), each a fixed amount created by one transaction and waiting to be spent by another, and it is this set, taken at a fixed block height, that we study. One observation is one UTXO, and its value is the integer number of satoshis it carries ($10^{8}$ satoshis per bitcoin). The unit is the protocol's own: an output is created by exactly one transaction, spent by at most one, and its value never changes in between. Spending it means satisfying the script that locks it, which in the common forms requires a signature valid under the private key for the public key that script names, so an output is controlled by whoever holds that key. We treat it purely as an accounting object. Nothing in what follows identifies who controls an output, or how many outputs one party controls.

\paragraph{Reconstruction.}
To obtain these sets we replay the active chain from block 1, reading raw blocks from a local Bitcoin Core node (version 30.2). Every output that Core treats as spendable is inserted, and every output that a later input consumes is removed. We follow Core's conventions in the three places where they matter: provably unspendable outputs, those with an \texttt{OP\_RETURN} prefix or a script longer than 10,000 bytes, never enter the set; the genesis coinbase is excluded; and the two duplicate coinbase transactions of 2010 overwrite their predecessors. To check the replay we compared it with an independent snapshot produced by Core's \texttt{dumptxoutset} at height 900,000, and the two agree exactly: the same block hash, the same 169,857,136 outputs, the same total of 19,874,782.09~BTC, the same 15,196 zero-value outputs, and the same SHA-256 hash of the sorted multiset of values. Along the way every input found a live output to spend, at every height.

\begin{figure}[t]
\centering
\includegraphics[width=\textwidth]{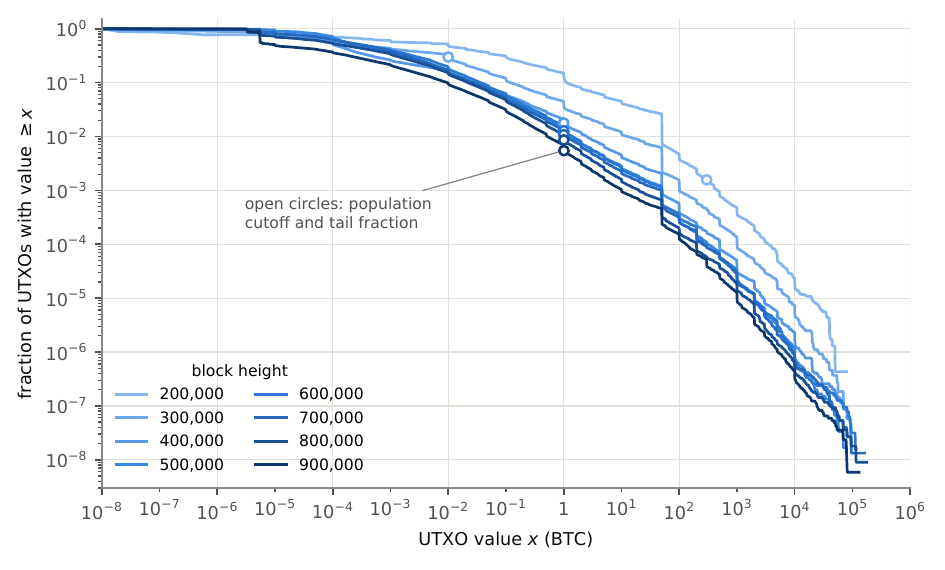}
\caption{Complementary cumulative distributions of UTXO values at the eight heights of Table~\ref{tab:manifest}, computed from every value in each population. Lighter lines are earlier heights. Open circles mark the cutoff and tail fraction selected on the population by the procedure of Section~\ref{sec:methods}: 1~BTC at six heights, 0.01~BTC at 300{,}000, and 299~BTC at 200{,}000. The body of the distribution moves toward small outputs over time while the tails above 1~BTC nearly coincide.}
\label{fig:ccdfpanel}
\end{figure}

\paragraph{Schedule.}
We fixed the heights before any fit: 200,000 to 900,000 in steps of 100,000, a panel that runs from September 2012 to June 2025 at intervals of about two years and cuts across market cycles rather than aligning with them (Table~\ref{tab:manifest}). Height 100,000 was also reconstructed, but with only 71,888 outputs it serves for description, not for testing. Figure~\ref{fig:ccdfpanel} already shows the pattern the rest of the paper examines: over thirteen years the body of the distribution slides toward ever smaller outputs, while the tails above 1~BTC lie almost on top of one another.

\begin{table}[t]
\centering
\caption{Snapshot manifest. Counts are from the validated replay; the supply in the UTXO set is the total value of live outputs, which is the cumulative block subsidy less provably unspendable outputs; the largest output is the largest single UTXO value; sampled outputs are the transaction-sampled subset used for the bootstrap (Section~\ref{sec:methods}).}
\label{tab:manifest}
\small
\begin{tabular}{rlrrrrr}
\toprule
Height & Date & UTXOs & Zero-valued & \shortstack[r]{Supply in the\\UTXO set (BTC)} & \shortstack[r]{Largest output\\(BTC)} & \shortstack[r]{Sampled\\outputs} \\
\midrule
200,000 & 2012-09-22 &   2,318,056 &  1,370 &  9,999,890 &  79,956 & 1,000,111 \\
300,000 & 2014-05-10 &  10,852,334 &  3,080 & 12,749,865 &  79,956 & 1,022,947 \\
400,000 & 2016-02-25 &  34,820,275 &  5,558 & 15,249,861 &  94,772 &   993,714 \\
500,000 & 2017-12-18 &  59,949,466 &  5,582 & 16,749,849 &  79,956 &   469,976 \\
600,000 & 2019-10-19 &  63,389,760 & 14,750 & 17,999,830 & 117,305 &   487,234 \\
700,000 & 2021-09-11 &  75,262,612 & 14,826 & 18,812,300 & 163,011 &   580,925 \\
800,000 & 2023-07-24 & 111,535,121 & 14,889 & 19,437,287 & 178,010 &   870,786 \\
900,000 & 2025-06-06 & 169,857,136 & 15,196 & 19,874,782 & 130,010 & 1,323,117 \\
\bottomrule
\end{tabular}
\end{table}

\paragraph{Sampling.}
The fits and tests of Section~\ref{sec:results} use a sample drawn by transaction; descriptive estimates over every value in the population appear in Appendix~\ref{app:supp}. The sample is a computational necessity, since the bootstrap refits the whole procedure thousands of times, which is out of reach on populations of up to 170~million outputs. A transaction enters the sample when the first two bytes of its identifier, read as an integer, fall below a height-specific threshold, and all its live outputs come with it. Selecting whole transactions matters: the outputs of one transaction are not independent draws, and the resampling scheme of Section~\ref{sec:methods} relies on keeping them together.

At the early heights the thresholds fix the sample size rather than the sampling rate, giving about one million outputs at each of the heights 200,000 to 400,000, between 2.9\% and 43\% of the population. The constant rate used from 500,000 onward would have left only 17,620 outputs at 200,000 (Table~\ref{tab:dense} sets the two samples side by side); from 500,000 onward that rate, 0.78\% of transactions, already yields samples of the same order. All the thresholds were fixed in the protocol from the population counts, not from any fit.

\paragraph{Exclusions and features of the data.}
The primary analysis applies no filter for dust, value, script type or age; zero-value outputs lie outside the positive support and are reported separately. Two features of the data shape what follows. The first is that round values are common. Exact multiples of 0.01~BTC make up between 17\% and 70\% of the tails fitted in Section~\ref{sec:results}, and at height 200,000, 12\% of the 349,000 outputs above 1~BTC are worth exactly 50~BTC: unspent block rewards.\footnote{Block rewards were 50~BTC before block 210,000 (November 2012). In our samples 92 to 95\% of the outputs worth exactly 50~BTC carry the coinbase flag, all were created before block 210,000, and their median creation height lies in the chain's first year, so most are rewards mined in 2009 and 2010 that have never been spent. A large share of these coins is commonly attributed to the earliest miners, including Bitcoin's creator; the ledger itself does not identify their owner. The remainder are ordinary payments of exactly 50~BTC.} We keep these atoms in the primary analysis and test their influence in Section~\ref{sec:results}.

The second is that the body of the distribution has moved steadily toward small outputs. The median positive output was 0.01~BTC in 2012 and 1,000 satoshis in 2025, by which time 78\% of all outputs held less than 0.001~BTC. By count, then, the tail we analyse is a thin slice of the ledger, the top 0.5\% to 4\% of outputs at the later heights, even though it holds most of the coins.

\paragraph{Flow window.}
The live stock is what survives of everything ever created, which raises the question of whether its shape is already present in the outputs being created. For each height the replay therefore also records the value of every output created, and of every output spent, in the 10,000 blocks ending at that height, about ten weeks, together with the live total before the window. Created minus spent reproduces the change in the live total to the satoshi at every height, which confirms that the flow records are complete; Section~\ref{sec:results} compares the stock with this flow. Each sampled output also carries its creation height, a coinbase flag and a script type, used only in the robustness analyses.

\section{Methods}
\label{sec:methods}

\paragraph{The protocol.}
We follow the four steps of \citet{Clauset2009}. The first chooses the lower cutoff $x_{\min}$ above which a power law is fitted\footnote{``Above the cutoff'' means larger in value. Figures~\ref{fig:ccdf900k} and~\ref{fig:ccdfpanel} plot complementary cumulative distributions on logarithmic axes: at each value on the horizontal axis the curve gives the share of outputs at least that large. The fitted tail is everything to the right of $x_{\min}$, marked by open circles in Figure~\ref{fig:ccdfpanel}, and the body set aside lies to its left.} by minimising the Kolmogorov--Smirnov (KS) distance between the empirical distribution of the values at or above the cutoff and the fitted law. The second estimates the exponent $\alpha$ of $p(x) \propto x^{-\alpha}$ by maximum likelihood on those values. The third asks whether the fit is good enough to be believed: a semiparametric bootstrap generates synthetic data sets from the fitted power law above the cutoff and from the empirical distribution below it, refits each with its own cutoff and exponent, and takes as $p$ the fraction of synthetic KS distances at least as large as the observed one. A value of $p$ below 0.1 rules the power law out; a larger value does not establish it. The fourth compares the power law by likelihood ratio with alternative distributions fitted to the same tail. \citet{Clauset2009} refer that comparison to Vuong's normalised statistic for non-nested pairs \citep{Vuong1989} and to a chi-squared test for the nested pair, the power law against the power law with exponential cutoff; neither reference is valid here, for reasons given below, and we report both only as descriptive values. We used the \texttt{powerlaw} package \citep{Alstott2014} for the fitting routines and wrote the bootstrap, the resampling and the alternative fits ourselves, since the package does not implement the bootstrap.

The target of these $p$-values needs a word, because the UTXO set at a height is a finite population, observed in full. The inferential question is nonetheless whether its empirical tail is compatible with a stochastic power-law model. The bootstrap therefore quantifies model-based uncertainty, not finite-population sampling uncertainty: how far a tail that a power law itself generated, on this transaction structure, would typically sit from its fitted law. The transaction sample of Section~\ref{sec:data} exists for computational reasons and changes nothing about that target.

\paragraph{Exact discrete support.}
Values are integers in satoshis, so we fit the discrete branch of the protocol. The power law is $p(x) = x^{-\alpha} / \zeta(\alpha, x_{\min})$ on integers $x \ge x_{\min}$, where the Hurwitz zeta function $\zeta(\alpha, x_{\min}) = \sum_{k \ge 0} (k + x_{\min})^{-\alpha}$ is the normalising constant \citep[\S 2]{Clauset2009}, and the KS distance is computed from its cumulative form, $\zeta(\alpha, x) / \zeta(\alpha, x_{\min})$ for the upper tail. At these values the lattice barely matters. The exponent is estimated with the closed-form approximation of \citet[eq.~3.7]{Clauset2009}, accurate to about 1\% for cutoffs above 6; every cutoff here exceeds $10^{5}$, and on the observed tails the approximation and the exact discrete maximum-likelihood estimate differ by less than $3 \times 10^{-8}$. For the same reason, with every retained cutoff above 680,000 satoshis, the alternatives are fitted with continuous tail likelihoods, an approximation we checked against discrete fits, which agree in sign in every replicate. Exact values also replace the binned data of the authors' earlier analysis: binned data require the bin-probability likelihood and binned KS statistic of \citet{VirkarClauset2014}, not exact-value estimators applied to bin representatives, and binning loses power.

\paragraph{The cutoff and its consequences.}
Empirical power laws, where they hold at all, hold only in the upper tail, which is why the fit starts at a cutoff. The body follows some other shape, and fitting the whole range biases the exponent, sharply when the cutoff is too low and mildly when it is too high \citep[Fig.~3]{Clauset2009}. The cutoff is therefore part of the hypothesis: the claim under test is ``a power law above $x_{\min}$'', and $x_{\min}$ is estimated from the data with its own uncertainty. Its position has two consequences worth keeping in view.

The first is that it defines what the claim is about. At the heights from 400,000 onward, the selected cutoffs leave a small share of outputs above them, holding nearly all the bitcoin in existence; the two early heights differ, and Section~\ref{sec:results} gives both shares at every height. What lies below is dust and change, for which no one claims a power law. For that reason every table reports the tail count, the tail fraction by outputs, and the share of total value above the cutoff.

The second is that, when the tail curves, the cutoff moves with the amount of data. With more observations the curvature becomes detectable at lower values, so the KS rule retreats to the shortest range over which a straight line still fits, and the exponent rises with it. The selected cutoff then becomes evidence about the shape of the tail, not a fixed feature of the data. We therefore report, beside the selected cutoff, the exponent estimated above fixed cutoffs of 0.1, 1, 10 and 100~BTC at every height, a profile of the local slope that does not depend on the search.

Candidate cutoffs must leave at least 2,500 values in the tail, which under the sampling of Section~\ref{sec:data} is 0.2 to 0.5\% of each sample. \citet{Clauset2009} found about a thousand values sufficient for reliable selection; the higher floor also keeps the test from failing to reject a power law on a tail too small to reject anything.

\paragraph{Several candidate cutoffs.}
Over millions of distinct values the KS objective has not one minimum but several near-tied ones, at cutoffs orders of magnitude apart and with different exponents, and settling on one would hide that ambiguity. We therefore screen 4,096 rank-spaced candidates among the eligible cutoffs, retain the sixteen best grid-local minima, and evaluate every observed value in the bracket around each. The observed fit is the best of these basins, and every bootstrap replicate reselects among them, so the uncertainty from the choice of cutoff enters the intervals; we report how often each basin is selected. Inference is conditional on the observed basin set, which we state as a limitation.

Two further analyses ask how much the procedure depends on what it is given. The first runs the same rule on the full population, with a two-level grid in place of the exhaustive refinement, which is infeasible on $10^{6}$ to $10^{7}$ distinct values. These point estimates carry no bootstrap, and so no $p$-value and no interval, since refitting the cutoff on every value of a population 2,500 times over is out of reach. Their role is comparison: they show what the rule selects when it can see every value, the limit of the sample-size dependence described above, and Appendix~\ref{app:basins} reports where they agree with the sample fits and where they do not. The second reruns the whole procedure with the floor at 1,000 and 5,000 and with 8 and 32 retained basins, at three heights (Appendix~\ref{app:sensitivity}). The verdict holds in every run, the basin count changes nothing reported, and the floor moves the selected cutoff only at 200,000, where the tail sits close to it.

\paragraph{Dependence.}
The outputs of one transaction are not independent draws: change outputs, batched payments and wallet templates tie their values together. Treating each output as a separate draw would overstate the effective sample size, which narrows the reference distribution of the KS distance and makes a goodness-of-fit test reject a true law too often \citep{GerlachAltmann2019}. Since our finding is a rejection, the null has to carry the dependence the data have, and the bootstrap therefore resamples whole transactions. Each of 2,500 replicates draws transactions with replacement from the sample of Section~\ref{sec:data}, keeps all their live outputs, and refits cutoff and exponent. It then forms its synthetic set by keeping the resample's values below the refitted cutoff and replacing those at or above it with draws from the refitted discrete power law, with the tail count held fixed. This departs from \citet[\S 4.1]{Clauset2009}, who draw the tail count binomially and the body with replacement; here the transaction resample already supplies both kinds of variability before the synthetic step. The same replicates give 95\% percentile intervals for the exponent and the tail fraction, reported beside the independent-observation standard error $(\hat\alpha - 1)/\sqrt{n_{\mathrm{tail}}}$ for contrast.

Resampling transactions still leaves one gap. Given the resampled transactions, the synthetic tail values are drawn independently, so dependence among the tail values of one transaction is not reproduced in the null. We address it in two ways. Appendix~\ref{app:calibration} measures the level of the complete procedure on synthetic samples with a true power-law tail and dependent values within transactions, and reports how much dependence each fitted tail actually contains. Because that dependence is material at some heights, Section~\ref{sec:robustness} also repeats the test against a null that reproduces it, drawing the synthetic tail values of one transaction through a Gaussian copula at the correlation measured on that height's observed tail, with everything else unchanged.

Dependence can also reach beyond the transaction. An exchange sweeping funds may emit several related transactions in one block, a mining pool pays out in patterns that recur block after block, and the fee regime and denomination habits of a given day are shared by every transaction confirmed in it. If such within-block dependence were strong, resampling transactions would still overstate the effective sample size. As a check, Section~\ref{sec:results} repeats the bootstrap with the creation block as the resampling unit, keeping every live output created in the same block together and changing nothing else; the v6 records carry the creation height for this purpose. Markedly wider block-level intervals would mean the transaction is too fine a unit, and we would report them instead; intervals that agree mean the dependence that matters at this scale is already captured. A second robustness check excludes coinbase outputs, which are a protocol artefact rather than a payment and concentrate at exactly 50~BTC in the early panel.

\paragraph{Alternatives.}
Four alternatives are fitted above the same cutoff: exponential, lognormal, stretched exponential, and power law with exponential cutoff (Table~\ref{tab:families}). The exponential is the light-tailed reference that any heavy-tailed sample rejects. The lognormal and the stretched exponential bend downward continuously on logarithmic axes, so their local slope steepens without limit; for a small shape parameter $\beta$ the stretched exponential approaches a power law with exponent $1 + \beta\lambda'$, which is why $\beta$ is weakly identified on these data. The cutoff power law keeps a power-law slope up to a scale $1/\lambda$ and then falls off.

Three of the four contain the power law itself, each in a limit: the cutoff power law at $\lambda = 0$, the stretched exponential as $\beta \to 0$ with $\beta\lambda'$ held fixed, and the lognormal as $\sigma \to \infty$ with $(\ln x_{\min} - \mu)/\sigma^2$ held fixed. Their maximised likelihood can therefore never fall below the power law's. That fact shapes everything that follows: for these three families the sign of the improvement is settled in advance, and only its size carries information.

Estimation and sampling of each family are described in Appendix~\ref{app:validation}. The stretched exponential is fitted by profile likelihood over $\beta$ with the scale solved analytically, and its estimator passed a pre-specified validation on synthetic data for parameter recovery, behaviour on true power-law tails, and convergence. The lognormal is maximised in a parameterisation that makes its power-law boundary reachable; an earlier parameterisation did not reach it, and the correction (amendment v9, Appendix~\ref{app:validation}) changed the reported comparisons.

\begin{table}[htbp]
\centering
\caption{The families compared. Densities are conditional on $x \ge x_{\min}$ with normalising constants omitted. The primary fit of the power law is discrete, normalised by the Hurwitz zeta function; the comparisons use continuous likelihoods above the cutoff, where the unit lattice is negligible. Tables report the stretched exponential's scale as $\lambda' = \lambda x_{\min}^{\beta}$ and the cutoff power law's rate as $q = \lambda x_{\min}$, so that $x_{\min}/q = 1/\lambda$ is the cutoff scale in satoshis.}
\label{tab:families}
\footnotesize\setlength{\tabcolsep}{5pt}
\begin{tabular}{@{}llp{0.37\textwidth}@{}}
\toprule
Family & Density, $x \ge x_{\min}$ & Behaviour on logarithmic axes \\
\midrule
Power law & $x^{-\alpha}$ & Straight line of slope $-\alpha$; the hypothesis under test. \\
Exponential & $e^{-\lambda x}$ & Falls faster than any power; the light-tailed reference. \\
Lognormal & $x^{-1}\exp\!\big(-(\ln x - \mu)^2 / 2\sigma^2\big)$ & A parabola in $\ln x$: the local slope steepens without limit. \\
Stretched exponential & $x^{\beta - 1} e^{-\lambda x^{\beta}}$ & Bends downward at a rate set by $\beta$; tends to a power law of exponent $1 + \beta\lambda'$ as $\beta \to 0$. \\
Power law with exponential cutoff & $x^{-\alpha} e^{-\lambda x}$ & Slope $-\alpha$ up to the scale $1/\lambda$, then an exponential fall; equals the power law when $\lambda = 0$. \\
\bottomrule
\end{tabular}
\end{table}
For each alternative we report the log-likelihood ratio per tail observation, positive when it favours the power law, with its 95\% percentile interval over the transaction replicates. These intervals describe how much the comparison moves when the sample is resampled by transaction; they are not the rule by which a model is chosen, for the reason given under decision rules below. Vuong's statistic and the chi-squared statistic for the nested pair are computed and kept in the deposited outputs, for comparison with studies that report them, but no conclusion here rests on either. Their reference distributions assume independent observations, which these are not; for the nested pair the null also places the parameter on the boundary of its range, where the chi-squared distribution with one degree of freedom is the wrong reference even under independence.

\paragraph{Effect sizes and the supply bound.}
With tails of thousands to hundreds of thousands of observations, a goodness-of-fit test would reject any idealised distribution sooner or later, so a rejection alone says little. What carries information is the size of the departure and how the alternatives compare. We report the observed KS distance relative to the median synthetic distance, the spread of the exponent across basins, and the scale of the cutoff power law with its interval. We also extrapolate each fitted pure power law over the population tail, to the number of outputs it predicts above the largest existing output and above the supply mined at that height, and to the total value it implies if truncated there. With an exponent below 2 these quantities grow without bound as the sample grows, so the extrapolation is a physical check on the fit as well as a statistical one.

\paragraph{Held-out comparison.}
Because three of the alternatives contain the power law, an improvement measured on the data that produced the fit is positive by construction, and its direction has to be judged on data the fit has not seen. We split the transactions at a height into five folds by an independent byte of the transaction identifier, which keeps whole transactions together and makes the split reproducible from the ledger. Each family is fitted on four folds, with the cutoff selected on those folds, and scored on the fifth by its log likelihood per held-out observation; the five folds are then pooled. We report each alternative's score minus the pure power law's, and the lognormal's minus the cutoff power law's for the pair the in-sample comparison leaves closest. The basin set is the one identified on the whole sample, so the comparison is partially conditional: the parameters are fitted out of sample, the candidate cutoffs are not.

\paragraph{Decision rules.}
``Rejected'' means $p < 0.1$ in the bootstrap; otherwise ``not rejected'', never ``confirmed''.

For the comparisons, containment raises a difficulty that the usual theory does not resolve. An improvement in fit is positive by construction, so it cannot be read as evidence on its own, and the boundary asymptotics that would normally calibrate it do not apply here. For the lognormal and the stretched exponential, a half chi-squared reference would hold for a fixed cutoff and independent observations, because the score at the boundary involves the log excess, whose moments are finite. For the power law with exponential cutoff no such argument is available at all: the score at $\lambda = 0$ involves the value itself, whose mean diverges when the exponent is below 2. And neither condition, a fixed cutoff or independent observations, holds in our setting.

We therefore use a measured reference and a predictive rule. The reference is the improvement obtained on synthetic samples whose tails are true power laws, drawn on a real transaction structure (Appendix~\ref{app:calibration}); improvements inside that range are reported as not distinguishable from what a power law itself produces. The rule is predictive: an alternative is called ``preferred'' at a height when it scores better than the power law on held-out transactions, an outcome that containment neither forces nor rules out. The improvement per observation is reported with its cluster interval, so that its size can be judged. When the interval of the direct difference between two alternatives covers zero at a height, we report that the data give no clear evidence of a difference between them there, which is absence of evidence, not evidence that they fit equally well. Rankings that differ between heights are reported as such and not aggregated into a single verdict.

``Heavy-tailed'' is used in the empirical sense: over the observed range the tail decays more slowly than an exponential. It is not a claim about the asymptotic tail, where a distribution counts as heavy-tailed only if no exponential ever bounds it. The two can disagree, and for the family that often fits best here they do: a power law with an exponential cutoff falls exponentially beyond its scale, so it is light-tailed asymptotically while behaving as a heavy tail over the range we observe, since its fitted scale lies above almost all of that range.

Finally, the scope of the rejection. It concerns the pure power law as a model of the tail above the selected cutoff, with at least 2,500 values, and it does not exclude approximate scaling over a finite range, whose extent the exponents at fixed cutoffs describe. Conversely, a power law that is not rejected on a tail holding a small fraction of outputs would be a statement about that tail, not about the distribution; the value share of each tail is reported so that the reader can judge what the tail contains.

\paragraph{Frozen protocol.}
The estimand, sample frame, snapshot schedule, cutoff rule, replicate count, alternatives and decision rules were fixed on 23 July 2026, before any inference on the panel. Every later change is a dated amendment, recorded with its reason and with a statement of what it leaves unchanged (Table~\ref{tab:amendments}). Before inference the amendments made the procedure feasible and its record complete: a grid search with retained basins replaced an exhaustive scan that proved infeasible, the alternatives moved to continuous tail likelihoods and gained the stretched exponential, and the early heights were sampled more densely. The four that came after the first results add descriptive analyses, among them the sensitivity analysis of Appendix~\ref{app:sensitivity}, and correct the fitting of the lognormal. Two changes in all were made after seeing a result and are marked post hoc; none changes the primary goodness-of-fit test. The full protocol and amendments are in Appendix~\ref{app:protocol}, the validation gates and audit trail in Appendix~\ref{app:validation}, and the code, environment lock and data manifests in Appendix~\ref{app:repro}.

\section{Results}
\label{sec:results}

\begin{table}[t]
\centering
\caption{Goodness of fit of the pure power law. Cutoff, tail count, tail fraction by outputs and share of total value above the cutoff are for the selected basin; the exponent interval is the 95\% percentile interval over 2,500 transaction-cluster replicates with basin reselection; KS is the observed Kolmogorov--Smirnov distance and KS ratio is that distance divided by the median distance of the synthetic samples, a scale that grows with the tail size for a fixed discrepancy; $p$ is the bootstrap $p$-value (1/2501 is the smallest reportable value). Heights 200,000, 300,000, 400,000 use the dense samples of Section 2; the others the 0.78\% samples, which coincide with the dense design from 500,000.}
\label{tab:gof}
\footnotesize\setlength{\tabcolsep}{4pt}
\begin{tabular}{rrrrrrlrrl}
\toprule
Height & \shortstack[r]{Cutoff\\(BTC)} & \shortstack[r]{Tail\\count} & \shortstack[r]{Tail\\(\% outputs)} & \shortstack[r]{Tail\\(\% value)} & $\hat\alpha$ & 95\% interval & \shortstack[r]{KS\\ratio} & $p$ \\
\midrule
200,000 & 169.769 & 2,743 & 0.27 & 53.3 & 2.069 & [1.933, 2.106] & 0.0755 & 6.3 & 1/2501 \\
300,000 & 0.010 & 313,071 & 30.63 & 99.9 & 1.463 & [1.449, 1.475] & 0.0483 & 42.1 & 1/2501 \\
400,000 & 1.021 & 17,054 & 1.72 & 95.5 & 1.622 & [1.611, 1.633] & 0.0439 & 9.1 & 1/2501 \\
500,000 & 1.002 & 5,546 & 1.18 & 93.3 & 1.675 & [1.526, 1.688] & 0.0364 & 11.1 & 1/2501 \\
600,000 & 1.010 & 6,202 & 1.27 & 92.5 & 1.706 & [1.538, 1.727] & 0.0355 & 4.9 & 1/2501 \\
700,000 & 1.018 & 6,242 & 1.07 & 92.5 & 1.682 & [1.633, 1.697] & 0.0349 & 6.8 & 1/2501 \\
800,000 & 1.007 & 7,633 & 0.88 & 90.4 & 1.761 & [1.680, 1.775] & 0.0338 & 9.2 & 1/2501 \\
900,000 & 0.083 & 45,780 & 3.46 & 98.0 & 1.695 & [1.687, 1.748] & 0.0310 & 10.2 & 1/2501 \\
\bottomrule
\end{tabular}
\end{table}

\begin{table}[t]
\centering
\caption{Alternatives on the selected tail. Log-likelihood ratio per tail observation, power law minus alternative (positive favours the power law), median and 95\% percentile interval over transaction-cluster replicates. The exponential is omitted: its median ratio is between $+1.1$ and $+2.8$ at every height with the interval far from zero. Held-out: five-fold transaction-clustered predictive log score per observation, lognormal minus truncated power law (positive favours the lognormal). }
\label{tab:alternatives}
\footnotesize\setlength{\tabcolsep}{3.5pt}
\begin{tabular}{rrlllr}
\toprule
Height & \shortstack[r]{Tail\\count} & Lognormal & Stretched exponential & Truncated power law & Held-out \\
\midrule
200,000 & 2,743 & -0.0014 [-0.0033, -0.0001] & -0.0014 [-0.0035, -0.0002] & -0.0017 [-0.0037, -0.0005] & -0.0003 \\
300,000 & 313,071 & -0.0020 [-0.0046, -0.0008] & -0.0023 [-0.0051, -0.0010] & -0.0047 [-0.0066, -0.0036] & -0.0028 \\
400,000 & 17,054 & -0.0100 [-0.0122, -0.0081] & -0.0114 [-0.0137, -0.0093] & -0.0133 [-0.0165, -0.0102] & -0.0030 \\
500,000 & 5,546 & -0.0036 [-0.0053, -0.0024] & -0.0039 [-0.0055, -0.0026] & -0.0033 [-0.0099, -0.0024] & +0.0006 \\
600,000 & 6,202 & -0.0023 [-0.0044, -0.0011] & -0.0026 [-0.0046, -0.0012] & -0.0048 [-0.0077, -0.0022] & -0.0033 \\
700,000 & 6,242 & -0.0021 [-0.0033, -0.0012] & -0.0022 [-0.0035, -0.0013] & -0.0007 [-0.0054, -0.0001] & +0.0053 \\
800,000 & 7,633 & -0.0026 [-0.0038, -0.0006] & -0.0027 [-0.0038, -0.0007] & -0.0021 [-0.0049, -0.0015] & +0.0010 \\
900,000 & 45,780 & -0.0017 [-0.0024, -0.0006] & -0.0017 [-0.0024, -0.0007] & -0.0011 [-0.0030, -0.0007] & +0.0007 \\
\bottomrule
\end{tabular}
\end{table}

\begin{table}[t]
\centering
\caption{Finite-supply check of the fitted pure power law. Population tail is the sampled tail count divided by the sampling fraction. Predictions extrapolate the fitted pure power law over the population tail; the bound is the total value of the live UTXO set at the height, which is the cumulative subsidy less provably unspendable outputs (160~BTC at height 200,000, 221~BTC at 900,000). The last column is the total value the fitted law implies for the tail if truncated at the mined supply, as a multiple of that supply.}
\label{tab:supply}
\footnotesize\setlength{\tabcolsep}{3pt}
\begin{tabular}{rrrrrrr}
\toprule
Height & \shortstack[r]{UTXO-set\\supply (M BTC)} & \shortstack[r]{Population\\tail} & \shortstack[r]{Largest existing\\output (BTC)} & \shortstack[r]{Predicted above\\largest existing} & \shortstack[r]{Predicted above\\mined supply} & \shortstack[r]{Tail value /\\mined supply} \\
\midrule
200,000 & 10.0 & 6,358 & 79,956 & 9 & 0.05 & 0.89 \\
300,000 & 12.7 & 3,321,336 & 79,956 & 2,127 & 204 & 175 \\
400,000 & 15.2 & 597,581 & 94,772 & 485 & 21 & 34 \\
500,000 & 16.7 & 707,440 & 79,956 & 345 & 9.3 & 19 \\
600,000 & 18.0 & 806,888 & 117,305 & 213 & 6.1 & 15 \\
700,000 & 18.8 & 808,692 & 163,011 & 229 & 9.0 & 19 \\
800,000 & 19.4 & 977,677 & 178,010 & 99 & 2.8 & 8.7 \\
900,000 & 19.9 & 5,877,076 & 130,010 & 288 & 8.7 & 20 \\
\bottomrule
\end{tabular}
\end{table}

\begin{table}[t]
\centering
\caption{Exponent of a power law fitted above fixed cutoffs, on the full population of values at each height (tail count in parentheses). Descriptive: no cutoff search, no uncertainty from reselection. It rises from 0.1 to 1~BTC and from 10 to 100~BTC at every height, is flat or slightly lower between 1 and 10~BTC at the later heights, and at a fixed cutoff changes little after 2014.}
\label{tab:profile}
\small
\begin{tabular}{rllll}
\toprule
Height & 0.1 BTC & 1 BTC & 10 BTC & 100 BTC \\
\midrule
200,000 & 1.442 (723,184) & 1.684 (349,077) & 1.736 (104,074) & 2.100 (10,826) \\
300,000 & 1.531 (1,304,826) & 1.687 (476,624) & 1.768 (130,747) & 2.281 (16,090) \\
400,000 & 1.559 (2,180,810) & 1.758 (717,514) & 1.808 (157,213) & 2.230 (18,829) \\
500,000 & 1.627 (3,211,512) & 1.796 (840,868) & 1.815 (167,039) & 2.183 (20,412) \\
600,000 & 1.653 (3,809,828) & 1.848 (952,246) & 1.801 (167,065) & 2.141 (22,033) \\
700,000 & 1.663 (3,927,860) & 1.832 (946,272) & 1.784 (167,991) & 2.098 (22,447) \\
800,000 & 1.704 (5,253,171) & 1.898 (1,129,561) & 1.820 (175,225) & 2.098 (21,038) \\
900,000 & 1.712 (5,241,689) & 1.883 (1,100,075) & 1.774 (173,106) & 2.161 (24,725) \\
\bottomrule
\end{tabular}
\end{table}

\begin{table}[t]
\centering
\caption{Improvement in total log-likelihood over the pure power law, in nats, for the fit to the observed tail at each height. The three families contain the power law in their closure, so the improvement cannot be negative and its size is what carries information. The benchmark comes from simulation: on 99 synthetic samples whose tails are true power laws drawn on a real transaction structure of comparable tail size, the lognormal's improvement has a 99th percentile of 2.2 nats and a maximum of 3.8 when the tail values are independent, and 4.6 and 4.8 when they are strongly dependent (Appendix~\ref{app:calibration}). Values at 200,000, and the cutoff power law at 700,000, lie inside that range; the rest lie far outside it.}
\label{tab:totalnats}
\footnotesize\setlength{\tabcolsep}{6pt}
\begin{tabular}{rrrrr}
\toprule
Height & Tail count & Lognormal & Stretched exponential & Cutoff power law \\
\midrule
200,000 & 2,743 & 3.1 & 3.2 & 4.1 \\
300,000 & 313,071 & 598.1 & 687.8 & 1,459.8 \\
400,000 & 17,054 & 169.8 & 193.7 & 225.8 \\
500,000 & 5,546 & 21.4 & 24.6 & 45.9 \\
600,000 & 6,202 & 13.1 & 14.9 & 33.5 \\
700,000 & 6,242 & 14.1 & 15.5 & 5.3 \\
800,000 & 7,633 & 9.0 & 10.3 & 28.7 \\
900,000 & 45,780 & 79.2 & 79.4 & 47.8 \\
\bottomrule
\end{tabular}
\end{table}

\subsection{The pure power law is rejected at every height}
\label{sec:rejected}

At every one of the eight heights the pure power law is rejected, and by the widest margin the design can express. The bootstrap $p$-value is $1/2501$ throughout: none of the 2,500 synthetic tails generated from the fitted power law, each refitted with its own cutoff and exponent, came as far from its fitted law as the data do (Table~\ref{tab:gof}).

The size of the departure is steady. The observed Kolmogorov--Smirnov distance falls from 0.076 at 200,000 to 0.031 at 900,000, and at seven heights it is five to eleven times the median synthetic distance. The exception is 300,000, where the ratio reaches 42 although the absolute distance there, 0.048, is ordinary for the panel. Two things inflate the ratio at that height: a very large tail, which makes the synthetic distances small, and dependence within transactions that the primary null does not reproduce. Against a null that does, the ratio falls to 21 (Section~\ref{sec:robustness}).

\begin{figure}[t]
\centering
\includegraphics[width=\textwidth]{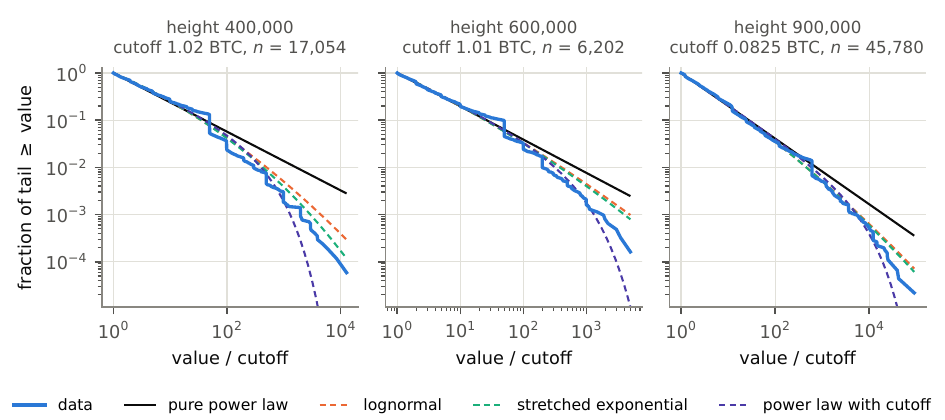}
\caption{Empirical distribution of the tail above the selected cutoff (blue) and the fitted models, at an early, a middle and a recent height; values are scaled by the cutoff. The pure power law is a straight line on these axes; the data leave it after one to two orders of magnitude in value. The three curved alternatives follow the data further; the power law with exponential cutoff bends most sharply.}
\label{fig:tailfits}
\end{figure}

Figure~\ref{fig:tailfits} shows where the departure lies. Above the cutoff the empirical tail follows the fitted straight line for one to two orders of magnitude in value and then falls below it, until at the largest values the shortfall reaches two orders of magnitude. The fitted exponents at the selected cutoffs run from 1.46 to 1.76 at seven heights; at 200,000 the exponent is 2.07, at a far higher cutoff (Section~\ref{sec:robustness}). Their 95\% intervals under transaction resampling are one to eight times as wide as the independent-observation intervals, and where they are wider, most of the width comes from replicates selecting a different cutoff basin (Section~\ref{sec:curved}).

These results replace those of the authors' earlier analysis of balances binned into nine logarithmic buckets \citep{BaqueroMenezes2026}, which reported exponents rising from 1.56 in 2013 to 2.89 in 2025 and did not reject the power law at the 2025 snapshot. Both features were artefacts of the binned support. With nine distinct values the KS statistic and the cutoff search have almost nothing to work with, and the 2025 non-rejection rested on a tail of 74 bin-level observations. On exact values the rejection holds at every height, including 900,000, and the exponents show no such rise (Table~\ref{tab:gof}).

\subsection{A heavy tail that curves, with no unique onset}
\label{sec:curved}

\begin{figure}[t]
\centering
\includegraphics[width=\textwidth]{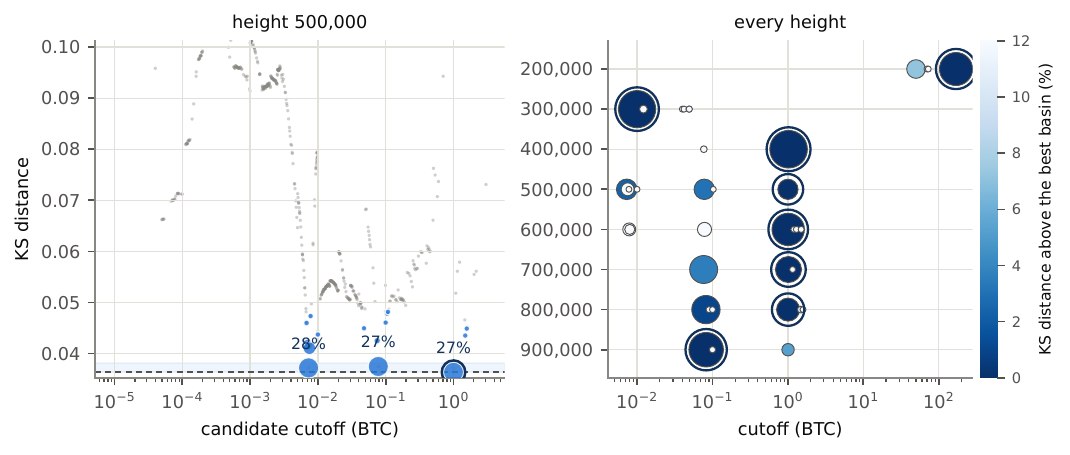}
\caption{Near-tied cutoff basins. Left: at height 500,000, the KS distance of every local minimum of the profile over the 4,096-point screen (grey) and the sixteen retained basins (blue, area proportional to the share of the 2,500 transaction-cluster replicates selecting each). The dashed line is the lowest KS distance, the band extends 5\% above it, and the ring marks the basin selected on the observed sample. Right: at every height, the basins that win at least one replicate, placed by cutoff and shaded by how far their KS distance lies above the best basin there; area is again the share of replicates and rings mark the observed selections. Where one basin takes most replicates the cutoff is well determined; where several share them, as at 500,000 and 800,000, the choice is unresolved over cutoffs orders of magnitude apart, and the bootstrap carries that uncertainty into the intervals.}
\label{fig:basins}
\end{figure}

The departure has a consistent shape, which three observations describe. The first is that the cutoff search does not find a single scale at which power-law behaviour begins. At 500,000, for example, three basins are near-tied: cutoffs of 1.0, 0.078 and 0.0073~BTC, whose KS distances differ by 0.0011 over tails that differ by a factor of seventeen (Figure~\ref{fig:basins}, left; Table~\ref{tab:basins}). Across cluster replicates each is chosen about a quarter of the time, and the exponent moves from 1.53 to 1.68 between them. At 900,000, by contrast, one basin dominates, taking 83\% of replicates, and the interval is correspondingly narrow. Figure~\ref{fig:basins} (right) and Appendix~\ref{app:supp} give the selection frequencies at every height.

\begin{figure}[t]
\centering
\includegraphics[width=\textwidth]{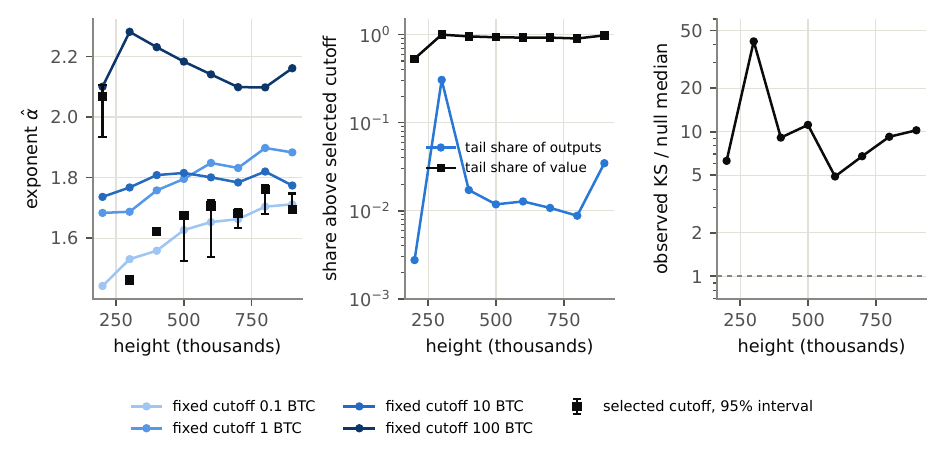}
\caption{The panel over time. Left: exponent at the selected cutoff with its 95\% cluster interval (black), and the exponent above fixed cutoffs of 0.1, 1, 10 and 100~BTC on the full population (blue). Middle: share of outputs and share of total value above the selected cutoff. Right: observed KS distance divided by the median synthetic distance; the dashed line marks equality.}
\label{fig:trajectories}
\end{figure}

The second is that the exponent depends on where the tail is cut. A power law fitted above a fixed cutoff summarises the whole tail above it, and on the full populations its exponent rises with the cutoff at every height, from 1.4 to 1.7 above 0.1~BTC to 2.1 to 2.3 above 100~BTC (Table~\ref{tab:profile}; Figure~\ref{fig:trajectories}, left). Between 1 and 10~BTC it is flat, or falls by up to 0.1 at heights 600,000 and later, where the 1~BTC denomination step sits at the lower cutoff. A power law has one exponent whatever the cutoff; an exponent that changes with it is the signature of a tail that curves on logarithmic axes, and it is what the goodness-of-fit test detects.

The third is that the selected cutoff depends on how much data the search can see. On the full populations the KS rule selects a cutoff within 1\% of 1~BTC at six of the eight heights, with exponents from 1.66 to 1.77 that agree with the sample-based fits, and at 300,000 population and sample both select 0.0100~BTC (Appendix~\ref{app:supp}). At 200,000, however, three samples of different sizes disagree, and in a telling direction. The 0.78\% sample of 17,600 values selects 0.47~BTC with exponent 1.53. The dense sample of one million values selects 170~BTC with exponent 2.07, on a tail of 2,743 outputs, with the 50~BTC atom as the runner-up basin in 22\% of replicates. The full population of 2.3~million values selects 299~BTC with exponent 2.13, on a tail of 3,638. With more data the curvature becomes visible at lower values and the rule retreats to the far tail: in the dense sample the eleven retained basins near 1~BTC have KS distances of 0.104, against 0.075 at 170~BTC (Section~\ref{sec:robustness} sets the samples side by side).

Figure~\ref{fig:trajectories} (middle) shows what the selected tails contain. At heights 400,000 and later they hold 0.9 to 3.5\% of outputs and 90 to 98\% of all bitcoin mined by then. The two early heights are the exceptions, the 0.01~BTC cutoff at 300,000 leaving 31\% of outputs above it and the 170~BTC cutoff at 200,000 only 0.27\%; Table~\ref{tab:gof} gives both shares at every height.

\subsection{Curved alternatives fit better, and which one fits best depends on the height}
\label{sec:alternatives}

\begin{figure}[t]
\centering
\includegraphics[width=\textwidth]{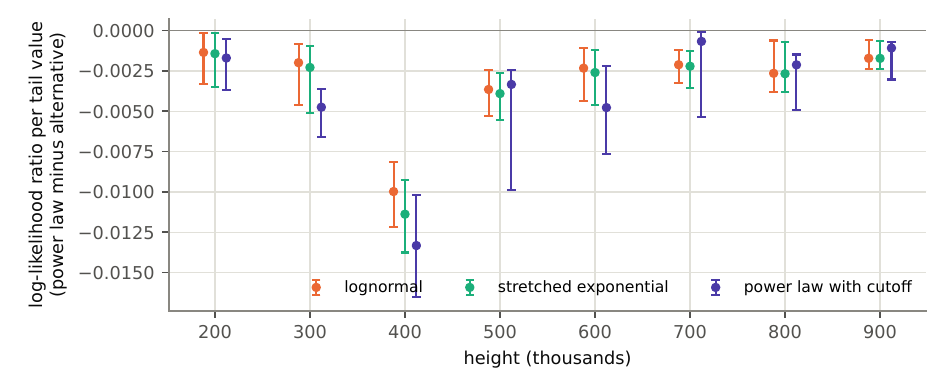}
\caption{Log-likelihood ratio per tail observation, power law minus alternative, with 95\% intervals over transaction-cluster replicates, for the three curved alternatives. Negative values favour the alternative. The exponential, at $+1$ to $+3$ nats per observation, is off this scale at every height.}
\label{fig:alternatives}
\end{figure}

Table~\ref{tab:alternatives} and Figure~\ref{fig:alternatives} compare the power law with the four alternatives on the same tails. The exponential loses by one to three nats per observation (natural-log units of likelihood) at every height, which settles that the tail is heavy. The other three contain the power law in their closure (Section~\ref{sec:methods}), so they cannot fit worse and only the size of their improvement matters. Each improves on the pure power law at every height, by 0.0014 to 0.0133 nats per observation, which in total is 3 nats at 200,000 and between 5 and 1{,}460 at the others (Table~\ref{tab:totalnats}).

Judging those totals requires knowing what a true power law yields, and simulation supplies it. On synthetic samples whose tails are true power laws, fitted the same way, the largest gain across the ninety-nine data sets with independent tail values was 3.8 nats for the lognormal, 3.6 for the stretched exponential and 4.7 for the cutoff power law; with dependent tail values the corresponding maxima run from 4.8 to 6.2 nats, reached at correlations of 0.90 and 1 (Table~\ref{tab:benchmark} in Appendix~\ref{app:calibration}). Four of the twenty-four observed gains sit inside the range for their own family, all three at 200,000 and the cutoff power law at 700,000, and are not evidence on their own. The other twenty, from 9 to 1{,}460 nats, are far outside it. These ranges were seen in particular simulations and have no known error rate, so exceeding one is informative and falling inside one does not show that a tail is a power law. The margins per observation are small because all three families differ from a power law only by a slow bend, which is what the data show.

Out of sample the improvement does not shrink, and the comparison is no longer one-sided. The lognormal predicts the held-out values better than the power law at every height, by 0.0014 to 0.0099 nats per observation, and the stretched exponential by 0.0015 to 0.0113; the cutoff power law predicts better at seven of the eight and loses at 700,000 (Table~\ref{tab:heldout} in Appendix~\ref{app:heldout}). A larger family pays out of sample for the parameter it gains in sample, so a positive held-out score is evidence that the shapes differ. It is not proof: chance can produce one, containment allows a tie, and the candidate cutoffs were identified on the whole sample, so the comparison is partially conditional (Section~\ref{sec:methods}). Even at 200,000, where the in-sample gain is inside the range a power law itself produces, all three alternatives predict better out of sample, in four folds of five.

No single one of the three fits best across the panel. The cutoff power law leads at 200,000, 300,000, 400,000 and 600,000, and the stretched exponential and the lognormal at the other four. The held-out comparison ranks the lognormal against the cutoff power law the same way at all eight heights, so the ordering is a property of each height and not of the fitting. Where the two are separated they are separated clearly: at 300,000 and 400,000 the interval of their direct difference excludes zero and the cutoff power law is ahead by 0.003 nats per observation (Appendix~\ref{app:paired}). At the other six the interval covers zero, and the data give no clear evidence of a difference between the two. The stretched exponential tracks the lognormal to within 0.0014 nats everywhere and is never behind it.

The cutoff power law also gives the tail a scale. Its exponential cutoff lies between 880 and 1{,}330~BTC at six of the eight heights, with cluster intervals spanning a few hundred to a few thousand bitcoin. The two exceptions are informative in different ways. At 700,000 the scale moves to 27{,}000~BTC because the sample happens to contain the 163{,}011~BTC output, the largest in the panel, and its interval spans three orders of magnitude. At 200,000 the selected cutoff is itself 170~BTC, above most of the range over which a bend could be seen, and the fitted scale of 62{,}000~BTC, with an interval from 19{,}000 to 163{,}000, means only that no cutoff is identified within the observed range. The scale is therefore reported with its interval, and the few dozen outputs above 10{,}000~BTC are better described from the full population than from a sample. Appendix~\ref{app:supp} gives the direct comparison between the curved alternatives, replicate by replicate, and the result of fitting each of them with a cutoff of its own.

\subsection{A pure power law with these exponents extrapolates far beyond the supply}
\label{sec:supply}

A different kind of evidence comes from the supply itself. Every unbounded distribution places some probability above a finite supply; what distinguishes a pure power law with exponent below 2 is how much. Its mean is infinite and its total is dominated by the largest members, since the typical size of the largest of $n$ draws grows like $n^{1/(\alpha-1)}$, faster than the sample itself. Table~\ref{tab:supply} extrapolates each fitted pure power law over the population tail. At the seven heights where the exponent is below 2 it predicts between 99 and 2{,}127 outputs larger than the largest output that exists, and between 3 and 204 outputs each larger than the entire supply held in the UTXO set at that height; truncated at that supply, the tail the fitted law implies is still worth 9 to 175 times the supply. At 200,000, the one place in the panel where the selected cutoff lies high enough for the fitted exponent to exceed 2, the same extrapolation stays within bounds: an exponent of 2.07 above 170~BTC predicts nine outputs above the largest existing one and a twentieth of an output above the supply.

The table is a diagnostic of extrapolation, not a proof of impossibility. What it shows is that the fitted power law cannot be taken literally at any scale of the tail, and that a description of these data has to bend, as the preferred alternatives do. They bend enough to stay near the supply: over the same population tail the lognormal and the stretched exponential predict fewer than one output above it at seven of the eight heights, and 14 and 5 respectively at 300,000, while the cutoff power law's prediction is far below one output at every height (Table~\ref{tab:supplyalt} in Appendix~\ref{app:supp}).

The supply argument does not, however, explain the rejection. The truncation it forces acts above about 1,000~BTC, which holds 0.04 to 0.3\% of the tail observations from 300,000 onward, so removing that slice moves the KS distance by at most that fraction, a hundredth to a tenth of what is observed. The largest deviation between the empirical and the fitted distribution sits well below the truncation scale: between 0.5 and 2~BTC at three heights, at 0.1~BTC at 300,000, and just below 50~BTC at three others, where the data fall short of the fitted law before the block-reward atom at 50~BTC. At 200,000, where the cutoff itself is 170~BTC, 18\% of the tail lies above 1,000~BTC and the exponent is above 2, so the supply argument does not apply. The physical and the statistical arguments are independent, and they point the same way.

\subsection{Stability from 2012 to 2025}
\label{sec:stability}

Across thirteen years the picture barely changes (Figure~\ref{fig:trajectories}). Above a fixed cutoff of 10~BTC the exponent stays within 0.08 of 1.79 throughout, and above 100~BTC it moves between 2.10 and 2.28 without trend; the exponents above the two lower fixed cutoffs drift upward by 0.2 to 0.3 across the panel as the body of the distribution moves toward small outputs (Table~\ref{tab:profile}). The exponent at the selected cutoff lies between 1.46 and 1.76 from 300,000 onward and at 2.07 at 200,000, where the cutoff sits at 170~BTC, and the differences among heights are mostly the cutoff moving rather than the tail changing shape. The KS ratio stays between five and eleven at seven heights, and the qualitative verdict of every test is the same at every height.

The composition of the tail tells the same story. From 300,000 onward the share of value above the selected cutoff stays above 90\%, while the share of outputs falls to about 1\% from 400,000 onward, and to 3.5\% at 900,000 where the sample selects the 0.08~BTC basin; that fall follows the body's collapse toward dust, not any change in the tail. Two cautions apply to reading the panel as a history. The eight snapshots share the outputs that survive from one to the next, so their agreement shows persistence over time; they are not eight independent replications. And the rows for 200,000 to 400,000 rest on the dense samples, which Section~\ref{sec:robustness} compares with the 0.78\% samples.

\subsection{Robustness}
\label{sec:robustness}

The checks in this section change one element of the analysis at a time: the sample, the resampling unit, the null, and the outputs included. The last comparison changes the object itself, from the stock of live outputs to the flow of newly created ones.

\paragraph{Dense samples at the early heights.}
Table~\ref{tab:dense} sets the 0.78\% sample and the dense sample side by side at heights 200,000 to 400,000. At 400,000 they agree in every respect: the same 1.02~BTC cutoff, exponents 0.002 apart, and an interval six times narrower on the larger tail. At 300,000 both select 0.01~BTC and the exponent moves from 1.475 to 1.463, while the KS ratio rises from 10 to 42, for the reasons given in Section~\ref{sec:rejected}.

At 200,000 the dense sample changes the fit, for the reason Section~\ref{sec:curved} describes: with more data the rule retreats from the 0.47~BTC basin of the small sample to 170~BTC, where the population estimate also lies. The verdict is unchanged at all three heights. We report the 200,000 row as the dense sample gives it, with the caveat that the dense tail sits close to the 2,500-observation floor and its exponent is the only one in the panel above 2; Appendix~\ref{app:sensitivity} shows how that row responds when the floor is moved.

\paragraph{Resampling by block.}
Table~\ref{tab:robust} repeats the bootstrap with the creation block as the resampling unit, which groups the same outputs into two to six times fewer, larger clusters. The two agree closely at every height: the endpoints of the exponent interval differ by at most 0.008, and by less than 0.002 at seven of the eight heights, the KS ratio by at most 0.2, and the $p$-value not at all. Dependence between transactions in the same block therefore adds little to the uncertainty that transaction clustering already captures. Because the check regroups an already transaction-sampled set of outputs, it does not show that no wider dependence exists.

\paragraph{A null that carries the dependence.}
The primary null draws the synthetic tail values of a transaction independently, so its reference is too narrow wherever those values move together. Table~\ref{tab:depnull} repeats the test at every height against the dependence-preserving null of Section~\ref{sec:methods}, at the correlation measured on each height's own tail (Table~\ref{tab:dependence}). At six heights the reference widens by 1 to 6\% and the KS ratio falls by at most a tenth. At 500,000 it widens by 29\%, and the ratio falls from 11.1 to 8.6. At 300,000 it widens by a factor of two, and the ratio falls from 42 to 21. That height is where the effect belongs: 58\% of its tail values share a transaction, against 3 to 18\% elsewhere, and one transaction holds 1{,}985 of them, so the independent null there understates the spread more than anywhere else in the panel. Under both nulls the $p$-value is the smallest the design can report at all eight heights. Half of the exceptional ratio at 300,000 therefore came from the null, and the rejection survives at every height.

This is a robustness analysis against one model of the dependence, not a general calibration. An exchangeable copula with a single correlation per height does not reproduce every feature of the joint distribution, and that correlation is estimated from the same data. The two flow windows whose ratios lie inside the range that dependence alone can produce, 2.9 at 300,000 and 2.7 at 600,000, were repeated the same way, and neither moves. The reference does not widen at either window, because those tails hold only 7\% and 12\% of their values in transactions with more than one, and the $p$-value at 600,000 falls from 0.0008 to the floor.

\paragraph{Coinbase outputs excluded.}
Removing coinbase outputs, 5.3\% of the sample at 200,000 and 1\% or less from 400,000, and re-identifying the basins leaves $p = 1/2501$ at every height (Table~\ref{tab:robust}), although the cutoff moves. At 200,000 it falls from 170 to 37~BTC, as the 50~BTC atom that anchored the runner-up basin disappears. At four of the five heights from 500,000 onward it rises from about 1~BTC to between 2.3 and 2.8~BTC, since the near-tied basins of Section~\ref{sec:curved} reorder when a few hundred outputs leave the sample. The exponent rises by 0.02 to 0.13 at the seven later heights and falls by 0.15 at 200,000, where the cutoff moves down. Where the cutoff rises the KS ratio falls, to between 2.8 and 5.2 from 500,000 onward, still with no synthetic tail as far from its fitted law as the data.

Round denominations are not what the test detects either. Removing every tail value that is an exact multiple of 0.01~BTC, 17 to 35\% of each tail from 300,000 onward, and refitting at the same cutoff leaves the KS distance at four to fifty times its null scale. This is a descriptive recomputation without a bootstrap, the null scale being the median synthetic distance adjusted for the smaller tail. The exponent rises by 0.05 to 0.2, because the round values sit low in each tail. At 200,000, where 70\% of the 170~BTC tail is round, too few values remain for the check to say anything.

\paragraph{Stock and flow.}
The last comparison changes the object. Tables~\ref{tab:flow} and \ref{tab:flowalt} apply the same fit and test to every output created in the 10,000 blocks ending at each height, spent or not, the flow of which the stock is the residue. The pure power law is rejected on the flow at every height, with $p = 1/2501$ at seven heights and $p = 0.0008$ at 600,000, and KS ratios between 2.7 and 23. The flow's cutoffs and exponents vary far more from height to height than the stock's, from 0.0009~BTC and 1.43 at 900,000 to 184~BTC and 2.21 at 300,000.

The alternatives also order differently. The cutoff power law improves on the pure law at every height. The lognormal and the stretched exponential improve at six, but at 200,000 and 600,000 their maximum likelihood is the power law itself, so they add nothing there; at 200,000 this holds in every one of the 2,500 replicates. A tail can be too heavy for a lognormal, and the flow at 200,000 is exactly that: the coefficient of variation of its log excesses is 1.03, above the value of one that separates the two regimes, while every stock tail lies between 0.88 and 0.96.

A heavy tail without a power law is therefore a property of the outputs being created as well as of those that survive, but the lognormal shape that describes the stock is not one the flow already has. The comparison sets a stock accumulated over the whole history against a window of 10,000 blocks, and the two samples overlap, so it cannot separate survival from changes in what was created, nor either from the difference in window.

\begin{table}[htbp]
\centering
\caption{The same test against two nulls. The independent null replaces the tail values of a resample by independent draws from the fitted power law, as the protocol specifies; the dependent null draws them through a Gaussian copula at $\rho$, the within-transaction correlation measured on that height's observed tail (Table~\ref{tab:dependence}). Everything else is identical, and at $\rho = 0$ the two coincide. The last column is the ratio of the two median synthetic distances, that is, how much reproducing the dependence widens the reference.}
\label{tab:depnull}
\footnotesize\setlength{\tabcolsep}{4pt}
\begin{tabular}{@{}rr rrl rrrl r@{}}
\toprule
 & & \multicolumn{3}{c}{Independent null} & \multicolumn{4}{c}{Dependent null} & \\
\cmidrule(lr){3-5}\cmidrule(lr){6-9}
Height & Observed KS & Median KS & Ratio & $p$ & $\rho$ & Median KS & Ratio & $p$ & Widening \\
\midrule
200,000 & 0.0755 & 0.0120 & 6.3 & 1/2501 & 0.66 & 0.0121 & 6.2 & 1/2501 & 1.01 \\
300,000 & 0.0483 & 0.0011 & 42.1 & 1/2501 & 0.57 & 0.0022 & 21.5 & 1/2501 & 1.96 \\
400,000 & 0.0439 & 0.0048 & 9.1 & 1/2501 & 0.75 & 0.0050 & 8.9 & 1/2501 & 1.03 \\
500,000 & 0.0364 & 0.0033 & 11.1 & 1/2501 & 0.63 & 0.0042 & 8.6 & 1/2501 & 1.29 \\
600,000 & 0.0355 & 0.0072 & 4.9 & 1/2501 & 0.90 & 0.0077 & 4.6 & 1/2501 & 1.06 \\
700,000 & 0.0349 & 0.0052 & 6.8 & 1/2501 & 0.56 & 0.0054 & 6.5 & 1/2501 & 1.05 \\
800,000 & 0.0338 & 0.0037 & 9.2 & 1/2501 & 0.50 & 0.0038 & 8.9 & 1/2501 & 1.04 \\
900,000 & 0.0310 & 0.0030 & 10.2 & 1/2501 & 0.56 & 0.0031 & 9.8 & 1/2501 & 1.04 \\
\bottomrule
\end{tabular}
\end{table}

\begin{table}[htbp]
\centering
\caption{Early heights on the 0.78\% transaction sample and on the dense sample of Section~\ref{sec:data} (43\%, 9.4\% and 2.9\% of transactions), labelled 0.78\% and dense. Basins are identified on each sample; intervals are 95\% percentile intervals over 2,500 transaction-cluster replicates with basin reselection.}
\label{tab:dense}
\footnotesize\setlength{\tabcolsep}{3pt}
\begin{tabular}{@{}rlrrrrrlrr@{}}
\toprule
Height & Sample & Outputs & Cutoff (BTC) & Tail $n$ & Tail \% & $\hat\alpha$ & 95\% interval & KS ratio & $p$ \\
\midrule
200,000 & 0.78\% & 17,620 & 0.467 & 3,417 & 19.39 & 1.532 & [1.378, 1.548] & 9.9 & 1/2501 \\
200,000 & dense & 999,266 & 169.8 & 2,743 & 0.27 & 2.069 & [1.933, 2.106] & 6.3 & 1/2501 \\
\addlinespace
300,000 & 0.78\% & 95,683 & 0.0100 & 27,769 & 29.02 & 1.475 & [1.432, 1.504] & 10.5 & 1/2501 \\
300,000 & dense & 1,022,225 & 0.0100 & 313,071 & 30.63 & 1.463 & [1.449, 1.475] & 42.1 & 1/2501 \\
\addlinespace
400,000 & 0.78\% & 270,183 & 1.022 & 4,648 & 1.72 & 1.620 & [1.523, 1.640] & 5.8 & 1/2501 \\
400,000 & dense & 993,220 & 1.021 & 17,054 & 1.72 & 1.622 & [1.611, 1.633] & 9.1 & 1/2501 \\
\bottomrule
\end{tabular}
\end{table}

\begin{table}[htbp]
\centering
\caption{Two robustness checks on the stock. Block level: the same sample and basins resampled by creation block instead of by transaction; the exponent interval is shown, and the KS ratio and $p$ (not shown) equal the transaction-level values of Table~\ref{tab:gof} at every height run. Coinbase excluded: coinbase outputs removed from the sample and basins re-identified; the share of outputs removed, the selected cutoff, the exponent with its interval, KS ratio and $p$.}
\label{tab:robust}
\footnotesize\setlength{\tabcolsep}{3pt}
\begin{tabular}{@{}rll rrrlrr@{}}
\toprule
 & Transaction & Block level & \multicolumn{6}{l}{Coinbase excluded} \\
\cmidrule(lr){4-9}
Height & 95\% interval & 95\% interval & Removed \% & Cutoff (BTC) & $\hat\alpha$ & 95\% interval & KS ratio & $p$ \\
\midrule
200,000 & [1.933, 2.106] & [1.932, 2.114] & 5.29 & 36.7 & 1.924 & [1.768, 1.944] & 9.9 & 1/2501 \\
300,000 & [1.449, 1.475] & [1.447, 1.475] & 2.58 & 0.0100 & 1.481 & [1.467, 1.496] & 45.6 & 1/2501 \\
400,000 & [1.611, 1.633] & [1.611, 1.633] & 0.98 & 1.000 & 1.719 & [1.706, 1.767] & 7.8 & 1/2501 \\
500,000 & [1.526, 1.688] & [1.526, 1.688] & 0.37 & 2.826 & 1.809 & [1.533, 1.830] & 5.2 & 1/2501 \\
600,000 & [1.538, 1.727] & [1.537, 1.728] & 0.31 & 1.002 & 1.790 & [1.542, 1.822] & 4.7 & 1/2501 \\
700,000 & [1.633, 1.697] & [1.633, 1.698] & 0.24 & 2.286 & 1.749 & [1.716, 1.778] & 2.8 & 1/2501 \\
800,000 & [1.680, 1.775] & [1.680, 1.774] & 0.16 & 2.496 & 1.842 & [1.696, 1.877] & 3.7 & 1/2501 \\
900,000 & [1.687, 1.748] & [1.687, 1.749] & 0.13 & 2.421 & 1.813 & [1.705, 1.838] & 3.2 & 1/2501 \\
\bottomrule
\end{tabular}
\end{table}

\begin{table}[htbp]
\centering
\caption{The flow of newly created outputs: every output created in the 10,000 blocks ending at the height, spent or not, sampled by transaction as the stock is. Fit and test as in Table~\ref{tab:gof}.}
\label{tab:flow}
\footnotesize\setlength{\tabcolsep}{3.5pt}
\begin{tabular}{@{}rrrrrrlrr@{}}
\toprule
Height & Outputs & Cutoff (BTC) & Tail $n$ & Tail \% & $\hat\alpha$ & 95\% interval & KS ratio & $p$ \\
\midrule
200,000 & 2,081,726 & 51.0 & 78,861 & 3.79 & 1.599 & [1.595, 1.604] & 23.0 & 1/2501 \\
300,000 & 1,126,873 & 183.8 & 3,653 & 0.32 & 2.206 & [1.815, 2.243] & 2.9 & 1/2501 \\
400,000 & 1,004,139 & 1.549 & 126,100 & 12.56 & 1.571 & [1.567, 1.574] & 12.4 & 1/2501 \\
500,000 & 446,039 & 11.6 & 13,801 & 3.09 & 1.862 & [1.841, 1.877] & 6.3 & 1/2501 \\
600,000 & 376,101 & 19.8 & 3,231 & 0.86 & 1.915 & [1.398, 1.979] & 2.7 & 0.001 \\
700,000 & 378,001 & 0.0050 & 178,233 & 47.15 & 1.464 & [1.460, 1.479] & 11.3 & 1/2501 \\
800,000 & 635,949 & 0.0060 & 167,305 & 26.31 & 1.468 & [1.462, 1.476] & 8.5 & 1/2501 \\
900,000 & 527,945 & 0.0009 & 251,444 & 47.63 & 1.428 & [1.423, 1.434] & 14.7 & 1/2501 \\
\bottomrule
\end{tabular}
\end{table}

\begin{table}[htbp]
\centering
\caption{Alternatives on the flow tails of Table~\ref{tab:flow}: log-likelihood ratio per tail observation, power law minus alternative, median and 95\% percentile interval over transaction-cluster replicates. The families contain the power law in their closure, so the ratio cannot be positive; a value of zero means the fitted alternative is the power-law limit itself, which happens in every replicate at 200,000 and in 740 of 2,500 at 600,000 for the lognormal and the stretched exponential.}
\label{tab:flowalt}
\footnotesize\setlength{\tabcolsep}{5pt}
\begin{tabular}{@{}rlll@{}}
\toprule
Height & Lognormal & Stretched exponential & Cutoff power law \\
\midrule
200,000 & +0.0000 [-0.0000, +0.0000] & +0.0000 [-0.0000, +0.0000] & -0.0053 [-0.0056, -0.0050] \\
300,000 & -0.0065 [-0.0098, -0.0010] & -0.0072 [-0.0106, -0.0011] & -0.0100 [-0.0134, -0.0041] \\
400,000 & -0.0069 [-0.0076, -0.0062] & -0.0073 [-0.0080, -0.0066] & -0.0088 [-0.0093, -0.0083] \\
500,000 & -0.0041 [-0.0119, -0.0028] & -0.0045 [-0.0130, -0.0031] & -0.0062 [-0.0139, -0.0045] \\
600,000 & -0.0001 [-0.0154, +0.0000] & -0.0001 [-0.0174, +0.0000] & -0.0003 [-0.0120, -0.0001] \\
700,000 & -0.0023 [-0.0030, -0.0007] & -0.0025 [-0.0034, -0.0008] & -0.0015 [-0.0020, -0.0010] \\
800,000 & -0.0014 [-0.0019, -0.0009] & -0.0015 [-0.0021, -0.0009] & -0.0036 [-0.0043, -0.0029] \\
900,000 & -0.0009 [-0.0015, -0.0005] & -0.0010 [-0.0016, -0.0005] & -0.0031 [-0.0036, -0.0026] \\
\bottomrule
\end{tabular}
\end{table}

\FloatBarrier

\section{Discussion}
\label{sec:discussion}

\paragraph{Interpretation of the result.}
The upper tail of live UTXO values is heavy, and it curves. A power law can be fitted above any fixed cutoff between 0.1 and 100~BTC, but its exponent depends on where the cutoff is placed, rising from about 1.5 to about 2.2 across that range, with a dip at the later heights (Table~\ref{tab:profile}). The Clauset--Shalizi--Newman test rejects the pure power law at every one of the eight heights, with observed KS distances five to forty times what the fitted law itself produces.

Three curved families, the lognormal, the stretched exponential and the power law with exponential cutoff, describe the tail better, by a small margin. They contain the power law in their closure, so the improvement is one-sided by construction and only its size is evidence: measured against what a true power law produces in simulation, twenty of the twenty-four gains exceed the range observed for their own family, while the three at 200,000 and the cutoff power law's at 700,000 do not. The three rank differently from height to height, and the direct comparison separates the lognormal from the cutoff power law at only two heights. Fitted with cutoffs of their own at two heights, the lognormal and the cutoff power law are rejected as well (Appendix~\ref{app:v7}). The two families tested that way fail on their own terms, which does not exclude families that were not tested. The alternatives are better than the power law, not right.

\paragraph{Sample size and the rejection.}
With tails of thousands to hundreds of thousands of values, a goodness-of-fit test would eventually reject any idealised model, so the $p$-values settle nothing on their own. What settles it is how large the departure is and how stable. The data leave the fitted line after one to two orders of magnitude and end two orders of magnitude below it (Figure~\ref{fig:tailfits}), the absolute KS distance stays between 0.031 and 0.076 across thirteen years, and the exponent moves by about half depending on where the tail is cut, where a power law would have one exponent whatever the cutoff.

The curved families gain more than chance supplies. Their improvement exceeds the range observed in the simulated power-law tails, except at 200,000 and, for the cutoff power law, at 700,000, and it survives on transactions held out of the fit. Neither is proof: a held-out gain can come from chance, and the candidate cutoffs were identified on the whole sample.

We measured the dependence within transactions at each height and repeated the test against a null that carries it. No verdict changes. The one exceptionally large effect size in the panel, the KS ratio of 42 at height 300,000, falls to 21 under that null, so half of it came from the independence assumption. That null matches one correlation per height, not every feature of the dependence, and the calibration behind it rests on one early transaction structure (Section~\ref{sec:robustness}; Appendix~\ref{app:calibration}).

More data does not straighten the tail; it moves the cutoff outward. At 200,000 the dense sample settles above 170~BTC, the only height in the panel where the exponent exceeds 2, and part of that movement is mechanical, since the rule needs 2,500 values above the cutoff.
\paragraph{The regime this belongs to.}
Earthquake sizes are the closest precedent. Seismic moment, the energy an earthquake releases, follows a power law with an exponent below 2, and because the moment a region can release over time is limited, the accepted description is the tapered Gutenberg--Richter law, which bends away from the power law at the largest events \citep{Kagan2002}. Bitcoin output values are in the same position: an exponent below 2 with a bounded total. In the catalogue of \citet{Clauset2009} every data set with an exponent clearly below 2 and a large tail ends ``with cut-off''; those that pass as pure power laws have exponents above 2 or small tails. Bitcoin output values sit in the first group. The supply bound of Section~\ref{sec:supply} says the same from the other side, and a description of the tail has to bend somewhere below the supply. Where it bends, and whether the bend is exponential or lognormal, the data do not decide.

\paragraph{Location of the coin supply.}
The fitted shape also says where the coin supply sits. Because the exponent is below 2, each order of magnitude of output size up to the cutoff scale holds about as much value as the one below it, or more. At every height, outputs between 10 and 10,000~BTC hold 70 to 82\% of the supply, and outputs above 1~BTC hold 91 to 99\% (Figure~\ref{fig:valuebands}, left; Table~\ref{tab:valuebands} in Appendix~\ref{app:supp}).

That profile moved slowly and in one direction. The 10 to 10,000~BTC band held 82\% of the supply in 2012 and 72\% in 2025, as value moved out of the 10 to 100~BTC class, where unspent 50~BTC rewards once dominated, and into the 100 to 1,000~BTC class (Table~\ref{tab:valuebands}). The number of outputs meanwhile grew from 2.3~million to 170~million while the supply doubled, and the median output fell to 1,000 satoshis (Figure~\ref{fig:valuebands}, right).

The stock is thus two populations: a growing mass of outputs that carry almost no value, and a tail whose value profile has changed only gradually since 2012. The bend in that tail is steady too. The scale at which the fitted cutoff power law bends lies between 880 and 1{,}330~BTC at six of the eight heights, a range it has kept in coins for eleven years, while its dollar equivalent rose from about four hundred thousand to a hundred and thirty million; at the other two heights it is not identified within the observed range. That scale is a parameter of a fitted family, not a measured operating limit of any participant, and no entity attribution is made here. Whatever produces the bend is nonetheless described by the same number of coins in 2014 and in 2025.

\begin{figure}[t]
\centering
\includegraphics[width=\textwidth]{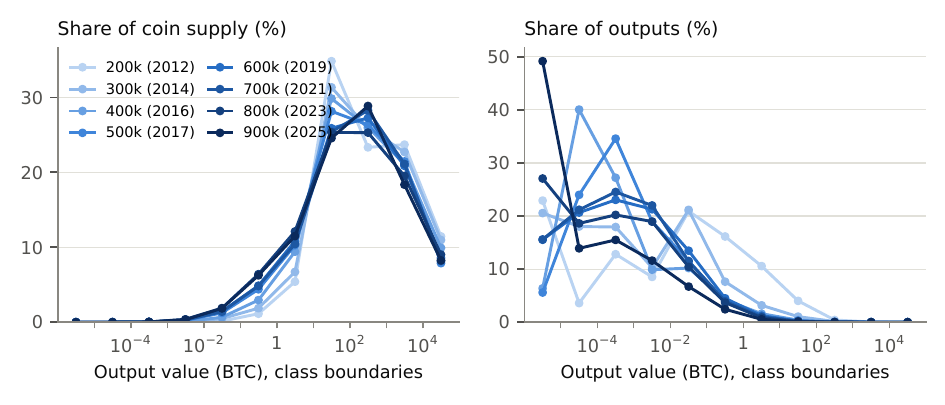}
\caption{Location of the live supply. Left: share of the live supply held in outputs of each size class, at the eight heights (lighter is earlier). Right: share of outputs in each class. Exact counts over every positive output; class boundaries in BTC.}
\label{fig:valuebands}
\end{figure}

\paragraph{Limits of the claim.}
The unit is the output (UTXO), not the holder. One entity owns many outputs and one output can be shared, so nothing here measures the distribution of wealth among people or entities; that question needs address clustering and belongs to a different literature \citep{Zhang2025,MakarovSchoar2021}. The stock is also survival-selected: an output enters it when created and leaves when spent, so its distribution is the flow filtered by holding behaviour. The flow of newly created outputs is heavy-tailed and not power-law either, but its shape differs from the stock's (Section~\ref{sec:robustness}), so the stock's shape is a property of what is kept, not only of what is made.

On the statistical side, the null of the primary bootstrap omits dependence within transactions. The calibration estimates what that omission does on one early structure, under the correlations simulated, and the copula null of Section~\ref{sec:robustness} measures it at each height under one model of the dependence; neither establishes the level of the procedure at every height under every possible form of dependence. Finally, the cross-sectional question answered here is separate from the time-domain claim that Bitcoin's price follows a power law in time. The two share a word and nothing else, and the time-domain question needs methods of its own, which we plan to develop in future work.

\paragraph{Mechanism.}
Nothing in the protocol identifies a generative model, and the paper does not propose one. The ingredients are known from the transaction-analysis literature (Section~\ref{sec:related}): block subsidies and wallet templates that create atoms at round values, coin selection and change that split values on a fixed grid, exchanges that batch and peel, and dust that accumulates below the fee floor. A model built from them would have a specific target to hit. It would have to produce a tail that is straight for one to two orders of magnitude and bends beyond, and let its value profile drift by ten points of the supply while the number of outputs grows seventy-fold. That is a more specific target than a power law.

\section{Related work}
\label{sec:related}

Power laws have been reported for several Bitcoin quantities, almost always from the slope of a log--log plot. Degree distributions of the address and user graphs came first \citep{Kondor2014,LischkeFabian2016,Maesa2018}. Later work with a formal test rejected the pure power law for monthly address degrees \citep{Liang2018}; other work reported exponents without a goodness-of-fit test \citep{Aspembitova2019}.

For values the record is thinner, and what exists is about balances rather than individual outputs. \citet{Kondor2014} fitted the tail of address balances at five dates and found, by inspection, that a stretched exponential described it better than a power law. \citet{Zhang2025} tested the balances of address clusters at one date with the uniformly most powerful unbiased test of \citet{Malevergne2011} and found a lognormal. Further from the value itself, \citet{Park2026} studied 63 recurring denominations and found that the number of such outputs held per address follows a geometric law; the denominations themselves are template amounts of wallet software. \citet{SornetteZhang2025} fitted power laws by regression to holding times and to age-dependent flows, and \citet{MakarovSchoar2021} measured the concentration of holdings among entities. None of these studies fits the value of individual outputs.

Output values have been described but not tested. \citet{LischkeFabian2016} tabulated the values of every output created in the first four years in seventeen bins and noted that most fall below one bitcoin. Two later studies described the UTXO set itself, \citet{DelgadoSegura2019} at two blocks of 2017 and 2018 and \citet{PerezSola2019} at nine heights from 100,000 to 500,000. About 98.5\% of outputs held less than one bitcoin. The most common amount was 1,000 satoshis, and powers of ten were over-represented. Neither fits a parametric form.

\citet{Li2019} is the closest parametric precedent. It fits monthly distributions of on-chain transfer sizes from 2009 to 2017 by least squares on the log-density, with a power law below a crossover size and an exponentially decayed power law above it. The tail exponents are about 0.27 and 1.75 in their cumulative convention, 1.27 and 2.75 in the density convention used here, and no goodness-of-fit test is reported. They analyse the flow of created values, not the stock of live ones. They also noticed the spike at 50~BTC in 2010, which our data show persisting as an atom of the live set.

\citet{Jourdan2018} plotted the output values of an entity-labelled subset and called the plot qualitatively power-law, and a survey \citep{Liu2021} credits \citet{Baumann2014} with an earlier visual observation of a power law in transaction values. Our own preprint \citep{BaqueroMenezes2026} applied the protocol to balances binned into nine buckets at yearly snapshots; Section~\ref{sec:results} explains why its rising exponents and its one non-rejection were artefacts of the binning.

The present study differs from all of these in four ways. Its object is the live stock of exact output values. Its method is a goodness-of-fit bootstrap that resamples transactions, a likelihood comparison among four families, and a calibration of the procedure on synthetic data. Its design covers eight heights spanning thirteen years, on a reconstruction validated against an independent snapshot. Its outcome is a rejection at every height, with the alternatives ranked.

The statistical background is the literature on testing power laws. \citet{Clauset2009} set the protocol, and our fits follow the conventions of its implementation by \citet{Gillespie2015}. \citet{StumpfPorter2012} argued that most claimed power laws come with neither a test nor a mechanism. \citet{BroidoClauset2019} applied the protocol to a thousand networks and found that few pass. \citet{Voitalov2019}, \citet{Holme2019} and \citet{Serafino2021} contested that conclusion, on the grounds that the definition of scale-free, the finite size of the samples and the choice of null shape the verdict.

Two points of that debate carry over to values. The first is separation. The pure power law is hard to tell from a truncated power law or a lognormal at moderate sample sizes, which is why tests built for that comparison \citep{Malevergne2011,BeeRiccaboniSchiavo2011,Corral2020,Bee2024} and estimators for the truncated power law \citep{DelucaCorral2013,Hanel2017} exist. Our held-out comparison and our direct comparison ask the same question, and find the two families separated at two of the eight heights and not at the rest. The second is dependence. \citet{GerlachAltmann2019} showed that correlated observations make goodness-of-fit tests reject a true law too often; our transaction-cluster resampling and the calibration in Appendix~\ref{app:calibration} measure this effect on the present data.

Size distributions in other economic systems set the expectation, and few of them are pure power laws. Firm sizes follow Zipf's law over several orders of magnitude \citep{Axtell2001}. City sizes are lognormal over most of their range with a Pareto tail of disputed extent \citep{Eeckhout2004,BeeRiccaboniSchiavo2013,Corral2020}. Income and wealth combine an exponential body with a Pareto tail \citep{Yakovenko2009}. In payment systems the values transferred are heavy-tailed \citep{Soramaki2007}. For interbank loan sizes \citet{Vandermarliere2015} found the configuration we find here: a truncated power law in the tail, and a lognormal or a stretched exponential over the full range. Seismology offers the closest analogue: the energy exponent is also below 2 and the total moment is conserved, and the accepted form is the tapered Gutenberg--Richter law \citep{Kagan2002}. Bitcoin output values belong with these bounded heavy tails, not with the clean power laws.

The mechanisms that shape output values are documented in the transaction-analysis literature. Block rewards create coinbase outputs at the subsidy: 50~BTC until block 210,000, halved every 210,000 blocks since, and 3.125~BTC from block 840,000 in April 2024. When fees are negligible, as they were in the early years, these outputs are exactly the subsidy. Coin-selection algorithms in wallet software decide which outputs are spent and whether a change output is created \citep{Erhardt2016}. Heuristics for identifying change exploit the fact that payments, not change, tend to be round amounts \citep{Meiklejohn2013,MoserNarayanan2022}. Exchanges batch withdrawals into long peel chains \citep{Kappos2022}, wallet templates fix recurring denominations \citep{Park2026}, and dust accumulates at the bottom of the range \citep{PerezSola2019}. These are the sources of the denomination atoms described in Section~\ref{sec:results} and of the dependence within transactions that the bootstrap has to respect.

\section{Conclusion}
\label{sec:conclusion}

Are the values of live Bitcoin UTXOs power-law distributed in their upper tail? On exact values at eight block heights from 2012 to 2025, the answer is no. The test is the Clauset--Shalizi--Newman protocol, adapted for discrete support, for several near-tied cutoffs, and for dependence within transactions.

The tail is heavy: its local exponent lies between 1.5 and 2.2, depending on where in the tail one looks. It also curves, and a pure power law is rejected at every height by a wide margin. A power law with an exponential cutoff near a thousand bitcoin and a lognormal both describe the tail better, without either being adequate on its own terms, and which of the two fits best changes from height to height. The shape changed only gradually through thirteen years of growth. At every height most of the coin supply sits in outputs between 10 and 10,000~BTC, the band this tail describes.

Later work can qualify this answer without reversing it. A future snapshot that a calibrated test found compatible with a pure power law would show that the pattern does not persist, not that it was absent here. A generative model of output values that predicted the bend would explain it. The related claim that Bitcoin's price follows a power law in time is a separate question, with its own data and its own pitfalls. We leave it to future work.

\bibliography{references}

\appendix
\section{Frozen protocol and amendments}
\label{app:protocol}

The analysis protocol was written and locked on 2026-07-23, before any inferential computation, together with a machine-readable copy and a manifest of SHA-256 hashes of every input and program. Every later change carries a new identifier, a date, the reason for the change and a statement of what it leaves unchanged. The record is the directory \texttt{protocol/} of the repository. Table~\ref{tab:amendments} lists the versions. Two changes were made after seeing a result and are marked post hoc; every other change was fixed before the computation it governs.

Versions v7 and v8 add descriptive analyses whose results appear only in the appendices and in one sentence each of the Discussion. Version v9 corrects a numerical error in the fitting of one alternative, which changes reported comparisons but no design, and adds a second null reported beside the primary one. Version v10 adds a sensitivity analysis of the cutoff procedure's fixed constants, reported in Appendix~\ref{app:sensitivity} and in one sentence of the Methods. No version changes the primary goodness-of-fit test.

\begin{table}[htbp]
\centering
\caption{Protocol versions. Dates are the days on which each document was locked. ``Post hoc'' marks changes made after a result was seen.}
\label{tab:amendments}
\footnotesize\setlength{\tabcolsep}{4pt}
\begin{tabular}{@{}p{0.14\textwidth}p{0.43\textwidth}p{0.36\textwidth}@{}}
\toprule
Version, date & Change & Reason \\
\midrule
v1, 2026-07-23 & Estimand (value in satoshis of every live UTXO), the panel of heights, the transaction-cluster sample, 2,500 replicates, a minimum tail of 2,500, cutoff search over every eligible distinct value, alternatives (exponential, lognormal, power law with exponential cutoff), and the decision rule $p<0.10$. & Pre-specification. \\
v2, 2026-07-24 & Cutoff search on 128 rank-spaced candidates plus 128 in the bracket of the minimum; a doubled-grid check at three heights required before production. & The search over every eligible value was impractical at height 900,000 even for a single fit. \\
v3, 2026-07-24 & Basin identification: a 4,096-point rank screen, the 16 best grid-local minima, exhaustive evaluation of every value in each bracket; every retained basin re-evaluated and the minimum reselected in every replicate. & The doubled-grid check found a second KS minimum at height 500,000 (0.73 against 7.8 million satoshis); a single grid can land in either. \\
v4, 2026-07-26 & Continuous tail likelihoods above the retained cutoffs for the alternatives; agreement in sign with the discrete comparison required before use. & Discrete optimisers needed about a week per height; the smallest retained cutoff, 679,997 satoshis, makes the unit lattice negligible. \\
v5, 2026-09-16 & Estimator and null documented as implemented (no computational change); per-replicate recording of both fits, with percentile intervals; stretched exponential added, with a validation gate; nested statistic for the power law against its cutoff version. & Recheck of the locked protocol against \citet{Clauset2009} before work resumed. \\
v5.1, 2026-09-16, post hoc & The recovery criterion of the stretched-exponential gate replaced by three criteria: information bound, product $\beta\lambda'$, tail CDF. & The criterion failed (5.94\% against 5\%) for a parameter that is weakly identified at this sample size; the estimator sits at the Fisher bound. Both gate runs are reported in Appendix~\ref{app:validation}. \\
v6, 2026-09-17 & Height-dependent sampling thresholds giving about one million outputs at heights 200,000 to 400,000; 32-byte output records with creation height, coinbase flag and script type; created and spent values per 10,000-block window; validation gates G1 to G5. & The 0.78\% sample pinned the cutoff at the early heights through the minimum tail; the robustness analyses need the added fields; the stock-versus-flow comparison needs the flow. \\
v6.1, 2026-09-18, post hoc, documentation only & Gate G5 evaluated against a cluster-aware standard deviation and kept advisory. & The height-300,000 sample exceeded the 2\% band by 1.5 cluster-aware standard deviations; G5 never invalidated a run. \\
v7, 2026-09-18 & Lognormal and cutoff power law treated as primary candidates, each with its own cutoff selection and bootstrap, at heights 500,000 and 900,000. Descriptive. & Whether any two-parameter family passes the test on its own terms. \\
v8, 2026-09-18 & Calibration of the complete procedure on synthetic power-law tails with within-transaction dependence; paired ratios between alternatives; supply extrapolation of the alternatives. Descriptive. & An independent review observed that the null does not reproduce dependence within a transaction. \\
v9, 2026-09-19, items 1 and 3 corrective & The conditional lognormal maximised in a parameterisation that contains its power-law boundary, and the stored comparisons corrected by exact replicate regeneration; a second bootstrap whose synthetic tail values carry the measured within-transaction correlation; that correlation measured on every fitted tail; a reference distribution for the likelihood improvement. & A second independent review found that the lognormal maximiser could not reach the boundary and returned likelihoods below the power law's, and that the claims made for the v8 calibration went beyond what it established. \\
v10, 2026-09-23 & Sensitivity of the cutoff procedure to its floor (1,000 and 5,000 against 2,500) and to its basin count (8 and 32 against 16) at heights 200,000, 500,000 and 900,000; the full procedure rerun with the primary seed. Descriptive. & The co-author's review asked whether the conclusion depends on either constant. \\
\bottomrule
\end{tabular}
\end{table}

Two manifests freeze the state of the inputs: one written on 2026-07-24 under v1, and one on 2026-09-17 covering protocol documents, source files, the replay outputs at nine heights, the extracted Core snapshot and the analysis outputs of v3 to v5 (126 hashed files). A third manifest, written when the analysis closed, adds the v6 replay manifests and gate reports and every analysis output of v6 to v9, including the per-replicate records.

\section{Validation and audit trail}
\label{app:validation}

\paragraph{Reconstruction against Bitcoin Core.}
The reconstruction matches Bitcoin Core exactly at height 900,000. We compared the replay's UTXO set with an independent dump produced by Core's \texttt{dumptxoutset}, rolled back to that height (block hash ending \texttt{968a}, 169,857,136 coins), and the two multisets of values agree: the same count, the same total of 1,987,478,209,074,494 satoshis and the same SHA-256 of the sorted value vector. The manifest of every height records the block hash, the UTXO count, the total value and the cumulative count and value of provably unspendable outputs.

\paragraph{The v6 replay.}
The v6 replay passed every gate that could invalidate it. The v6 program changes the sampling rule and the output record but not the UTXO-set logic; a regression run to heights 100,000 and 200,000 (19 minutes) preceded the full pass (12.99 hours), and Table~\ref{tab:gates} gives the gates of amendment v6 on the full pass. G1 to G4 pass at every height. G5, a sanity check on the sampling thresholds, is met at heights 200,000 and 400,000 and exceeded at 300,000 by 22,225 outputs against a 2\% band of 20,000. The band was set as if outputs were sampled individually, whereas sampling is by transaction: transactions at that height hold up to 2,220 live outputs, and the standard deviation of the sample size under transaction sampling is about 14,500, so the excess is 1.5 of these standard deviations. Amendment v6.1 records the recalibration, and the first gate report is retained beside the final one.

\begin{table}[htbp]
\centering
\caption{Gates of amendment v6 on the full replay pass.}
\label{tab:gates}
\footnotesize
\begin{tabular}{@{}lp{0.60\textwidth}p{0.27\textwidth}@{}}
\toprule
Gate & Criterion & Result \\
\midrule
G1 & Sorted SHA-256 of the value vector equals the validated v1 replay (and, at 900,000, the Core snapshot) & Pass at all nine heights \\
G2 & Zero failed removals of spent outputs & Pass \\
G3 & Sample of (txid prefix, index, value) identical to the v1 sample at 500,000 to 900,000 & Pass at all five heights \\
G4 & Created minus spent value per window equals the change in the live total, exact in satoshis & Pass at all nine heights and windows \\
G5 & Positive sampled outputs within 2\% of 1,000,000 at 200,000 to 400,000 & 999,266; 1,022,225; 993,220 \\
\bottomrule
\end{tabular}
\end{table}

\paragraph{Cutoff search.}
The doubled grid confirmed the selected cutoff at two of the three heights where it was required, and at the third it exposed the problem that the basin procedure now handles. Amendment v2 required the coarse-to-fine search to be repeated with a doubled grid before production. At heights 200,000 and 900,000 the two grids select the same cutoff to within a few candidates (46,737,974 satoshis at both grids at 200,000; 8,267,705 against 8,253,284 at 900,000, with the exponent unchanged to three decimals). At 500,000 the doubled grid selected a different local minimum, 728,752 against 7,796,430 satoshis, with KS distances 0.0372 and 0.0375. That observation led to amendment v3, under which every local minimum of the KS profile is retained and reselected in every replicate.

\paragraph{Exponent estimator.}
The closed-form estimator and the exact one agree far more closely than sampling error could detect. The estimator in force is the closed-form approximation of \citet[eq.~3.7]{Clauset2009} when its value lies in $(1.5, 3]$ and the exact discrete maximum-likelihood estimate otherwise. On the observed tails at heights 200,000, 500,000 and 900,000 the two differ by at most $3\times10^{-8}$, against independent-sample standard errors of 0.002 to 0.009.

\paragraph{Continuous tail likelihoods.}
Continuous likelihoods reproduce the sign of the discrete comparison in every replicate, which amendment v4 required before their production use. The production run of 2026-07-26 preceded the written record, which was made on 2026-09-16 from the artefacts that exist. At height 500,000, on the same cluster index and basin set, 171 discrete replicates (the abandoned v3 run) and 2,500 continuous replicates (v4) give the same sign in every replicate for all three alternatives. The per-observation medians of the log-likelihood ratio are $-0.00372$ against $-0.00365$ nats for the lognormal, $-0.00331$ against $-0.00334$ for the cutoff power law, and $+1.842$ against $+1.856$ for the exponential; per-basin medians agree to within 0.0002 nats for the two curved alternatives and within 4\% for the exponential.

\paragraph{Estimation and sampling of the alternatives.}
All alternatives are fitted on the scaled tail $r = x / x_{\min} \ge 1$ with continuous likelihoods (amendment v4). The power law's exponent has the closed form $\hat\alpha = 1 + n / \sum \ln r_i$ and the exponential's rate is the reciprocal of the mean excess over the cutoff. The lognormal is maximised over $(k, t)$ with $t \ge 0$ as described below, which reaches its power-law boundary; the earlier maximisation over $(\mu, \ln \sigma)$ inside a box, which did not, is superseded for the comparisons and survives only in the runs of amendment v7, whose fits are interior.

The stretched exponential is written on $r$ as $p(r) \propto r^{\beta - 1} \exp(-\lambda' (r^{\beta} - 1))$ with $\lambda' = \lambda x_{\min}^{\beta}$; for fixed $\beta$ the maximum-likelihood scale is $\hat\lambda' = n / \sum (r_i^{\beta} - 1)$ in closed form, and the profile likelihood is maximised over $\beta$ by evaluating a nine-point grid from 0.05 to 2 and then a bounded one-dimensional search in $\ln \beta$ within the bracket of the best grid point. The cutoff power law is written as $p(r) \propto r^{-\alpha} e^{-q r}$ with $q = \lambda x_{\min}$; its normalising integral $J(\alpha, q) = \int_1^\infty r^{-\alpha} e^{-q r}\,\mathrm{d}r$ is computed by adaptive quadrature in $\ln r$, and the likelihood is maximised numerically over $(\alpha, \ln q)$ with the same optimiser pair. For the bootstraps of amendment v7, synthetic lognormal tails are drawn by inverting the conditional distribution function, and synthetic cutoff-power-law tails by inverting a cumulative trapezoid integral of the density on a logarithmic grid of 40,001 points.

\paragraph{Stretched-exponential gate.}
The stretched exponential failed one criterion of its validation gate, and the diagnosis traced the failure to weak identification of one parameter, not to a faulty estimator. Amendment v5 required, before production use, that the estimator recover its parameters on 200 synthetic tails matched to the observed fit at height 900,000 (cutoff 8,252,000 satoshis, $n = 45{,}780$, $\beta = 0.0441$, $\lambda' = 14.82$), that it behave correctly on true power-law tails, and that it converge. Table~\ref{tab:segate} shows both runs. The original recovery criterion failed: the median absolute relative error of $\hat\beta$ was 5.94\% against a 5\% threshold. The diagnosis, made before any change, showed that $\log\hat\beta$ and $\log\hat\lambda'$ are correlated at $-0.998$ across draws, that the product $\beta\lambda'$, which sets the tail slope in the small-$\beta$ limit, is recovered to 0.5\%, that the fitted tail CDF is within sampling noise of the truth, and that the observed error equals the median error of an efficient estimator at the profile Fisher information. Amendment v5.1 replaced the criterion by three that test what the analysis uses, and the gate was re-run with a fresh seed.

\begin{table}[htbp]
\centering
\caption{Stretched-exponential validation gate at height 900,000, 200 synthetic tails per set. The first run under v5 (seed 20260916) and the re-run under v5.1 (seed 20260917).}
\label{tab:segate}
\footnotesize\setlength{\tabcolsep}{4pt}
\begin{tabular}{@{}p{0.17\textwidth}p{0.40\textwidth}p{0.27\textwidth}l@{}}
\toprule
Criterion & Rule & Result & Passed \\
\midrule
\multicolumn{4}{@{}l}{\emph{v5, first run}} \\
(a) recovery & median absolute relative error of $\hat\beta$ below 5\% & 5.94\% (95th percentile 16.8\%) & No \\
(b) null behaviour & median per-observation LLR on true power-law tails within 0.001 of zero & $1.3\times10^{-6}$ & Yes \\
(c) convergence & at least 99\% of fits converge & 100\% & Yes \\
\multicolumn{4}{@{}l}{\emph{v5.1, re-run}} \\
(a1) information bound & median absolute relative error of $\hat\beta$ at most $1.25\times0.6745\,\sigma_{\mathrm{rel}}$ & 5.25\% against 6.56\% ($\sigma_{\mathrm{rel}} = 7.78\%$) & Yes \\
(a2) product & median relative error of $\hat\beta\hat\lambda'$ below 5\% & 0.49\% (95th percentile 1.30\%) & Yes \\
(a3) tail CDF & median KS distance between true and fitted tail CDF below $0.8276/\sqrt{n}$ & 0.00143 against 0.00387 & Yes \\
(b) null behaviour & as above & $7.4\times10^{-7}$ & Yes \\
(c) convergence & as above & 100\% & Yes \\
\bottomrule
\end{tabular}
\end{table}

\paragraph{Correction of the lognormal maximiser.}
The first version of the lognormal fit could not reach the power law it contains, and in some fits it returned likelihoods that are impossible for the true maximum; amendment v9 corrected it. The comparison runner maximised the conditional lognormal likelihood over $(\mu, \log\sigma)$ inside a box. Writing $y = \log(x / x_{\min})$, that density is proportional to $\exp(-k y - t y^2/2)$ with $k = (\log x_{\min} - \mu)/\sigma^2$ and $t = 1/\sigma^2$, so the family is an exponential family in $(y, y^2/2)$ whose log-likelihood is concave in $(k, t)$, and the power law with exponent $1 + k$ is the boundary $t = 0$. That boundary lies at $\sigma \to \infty$ and $\mu \to -\infty$, outside any box on $\mu$. The maximiser therefore ran to the edge, and in ten of sixteen observed fits returned a likelihood below the power law's. Amendment v9 maximises over $(k, t)$ with $t \ge 0$ instead. The estimator was validated before use: it recovers the parameters of synthetic conditional lognormals, lands exactly on the boundary for synthetic Pareto tails and returns the power-law likelihood there, and over four hundred random tails never returns less than the power law's likelihood. The sign of the derivative in $t$ at the boundary is negative exactly when the coefficient of variation of the log excesses is below one, the statistic of the uniformly most powerful unbiased test of Pareto against lognormal, so the cases where the lognormal degenerates to the power law are identified exactly rather than by the optimiser's behaviour.

Each resample is a deterministic function of the seed and the replicate index, and the comparison runner draws nothing else, so the stored runs were corrected by regenerating every replicate and refitting the lognormal. The regeneration is verified on each of the 40,000 replicates by reproducing the recorded power-law and exponential likelihoods, which the correction does not touch, and the correction is refused if any replicate fails. In the corrected output none of the 40,000 replicates has an alternative below the power law. The corrected fits change the stock comparisons by at most 0.0004 nats per observation and the flow comparisons by up to 0.0058; three ratios that had favoured the power law, which is impossible for a family containing it, become zero or negative. The held-out comparison was re-run with the corrected fit and moved by at most 0.0003 nats. The runs of amendment v7 are unaffected: their fits are interior and none of their 5,000 replicates reaches the bound. The stretched exponential and the cutoff power law contain the power law in their closures too, at $\beta \to 0$ and $q \to 0$; where a bounded search returned less than the power-law likelihood the value is raised to that limit, which happened for the stretched exponential in the flow tails at 200,000 and 600,000 and nowhere else.

\paragraph{Reproduction of the v3 bootstrap under v5.}
The per-replicate recording added by amendment v5 left the v3 bootstrap exactly as it was, since it did not change the random-number stream. The v3 replicates at heights 200,000, 500,000 and 900,000 were regenerated from their seed, and acceptance required the selected cutoff of every one of the 2,500 replicates to match the v3 record; all three heights were accepted. In a 30-replicate check the synthetic KS distances were reproduced exactly.

\paragraph{The dependence-preserving null.}
The second null of amendment v9 behaves as intended in both directions. At $\rho = 0$ its synthetic distances are indistinguishable from those of the protocol's null at height 500,000, as they should be, since the two constructions then differ only in the route from a uniform variate to a value (two-sample KS distance 0.078 on 150 against 2,500 replicates, $p = 0.33$). At the correlations used, the synthetic tails it produces carry the intended dependence: measured with the estimator of Table~\ref{tab:dependence}, two replicates at height 600,000 give 0.86 and 0.94 against an intended 0.898, and two at 200,000 give 0.78 and 0.67 against 0.660, the value the run was given; the final estimator of Table~\ref{tab:dependence} puts that height at 0.65, a difference without consequence for the check. The spread is wider at 200,000 because the correlation there rests on few pairs: the observed tail contains 236 within-transaction pairs, 190 of them from a single transaction holding twenty tail outputs, and the two replicates measured happened to contain 36 and 37.

\paragraph{Consistency at the boundary.}
A family reported at its power-law limit must be used at that limit everywhere downstream, and a check now enforces it. The fitting code records the limit explicitly, and a check run after any change to it verifies, on an interior fit and on two fits at the boundary, that each family's pointwise log-likelihoods are finite, that they sum to the reported total, and that a family at its limit gives exactly the power law's density; the held-out scorer is exercised on the same tails. An earlier version of the pointwise code evaluated the lognormal through $(\mu, \sigma)$, which are infinite at the limit, and left the stretched exponential's finite parameters in place after raising its total, so the two disagreed by 3.7 nats on the flow tail at height 200,000. No result in this paper used those pointwise values: they enter only Vuong's statistic and the nested chi-squared, which the paper does not use, and the held-out scoring, whose forty training fits are all interior.

\paragraph{The within-transaction correlation.}
The correlation reported in Table~\ref{tab:dependence} and the one used to generate the dependent null differ by at most 0.009. The reported correlation is computed from van der Waerden scores with tied values sharing an average rank, so that the many repeated denominations do not make the value depend on the order of the array. The correlations that generated the dependent null were computed before that change, with ties broken by position, so each null is matched to a correlation within 0.009 of the one stated.

\paragraph{Calibration under dependence.}
The v8 calibration (Appendix~\ref{app:calibration}) tests the complete procedure, from basin identification to the bootstrap $p$-value, on synthetic samples whose tail is a true power law and whose outputs within a transaction are dependent. It is a check of the procedure's level, run after the main results were known; its design was fixed before it ran.

\section{Reproducibility}
\label{app:repro}

\paragraph{Code.}
All programs are in one versioned repository with dated commits from 2026-09-16; the July 2026 state, when the pipeline and the protocol were first written, is preserved as a tagged first commit whose files are byte-identical to the originals. The replay is a C program that reads blocks from a local Bitcoin Core node (version 30.2) over RPC and maintains the UTXO set in a 14~GiB open-addressing table with 28-byte slots; the analysis is in Python 3.12.9 with numpy 2.5.3, scipy 1.18.1, pyarrow 25.0.1 and the \texttt{powerlaw} package 2.0.0 \citep{Alstott2014}, with versions locked in the environment file. Runs were made on an Apple M3 with 24~GiB of memory and an external solid-state disk.

\paragraph{Data.}
For each height the replay writes the exact vector of positive UTXO values (unsigned 64-bit integers, 17~MiB at height 200,000 to 1,295~MiB at 900,000), the transaction-cluster sample with its cluster index, and a manifest with the block hash, the UTXO count, the total value and the sampling rule. The v6 pass adds the 32-byte output records and the per-window created and spent values (23~GB in all). The Core snapshot used for validation (10~GB) is reproducible from any archival node with \texttt{dumptxoutset} at height 900,000 and is not redistributed. The value vectors, the cluster samples and indices, the manifests, the created and spent values of the flow windows, the per-replicate records and every analysis output are deposited on Zenodo (DOI 10.5281/zenodo.22922174, which resolves to the latest version; about 10~GB), together with the repository as a full-history archive; the small outputs (JSON summaries, per-replicate records) are also mirrored inside the repository. Code is released under the MIT licence, data and outputs under CC~BY~4.0.

\paragraph{Seeds and records.}
Every bootstrap replicate $r$ uses a generator seeded from the pair (run seed, $r$), so any replicate can be regenerated alone. The goodness-of-fit runs use seed 20260724, the comparison runs 20260726, the v7 runs 20260918, and the v8 calibration a seed base of 20260918 for the synthetic data with 20260724 for the bootstrap on each synthetic sample. Each run appends one record per replicate to a checkpoint file, holding the selected cutoff, exponent, KS distance and tail count of the cluster resample and of its synthetic counterpart, and for the comparison runs the fitted parameters and total log-likelihood of every model. Every interval, $p$-value and ratio in the paper can be recomputed from these records without re-running a fit.

\paragraph{Cost.}
With 2,500 replicates, a goodness-of-fit run takes from 1.5 minutes at height 200,000 to 130 minutes at 800,000 on one core under the v1 sample; the v6 samples at 200,000 to 400,000 and the flow windows take one to five hours each. The v6 replay to height 900,000 took 13 hours. The v8 calibration takes about one minute per synthetic sample with 1,000 replicates.

\paragraph{Manifests.}
The frozen manifests (Appendix~\ref{app:protocol}) hold SHA-256 hashes of the protocol documents, the source files, the replay outputs at every height, the extracted Core snapshot and the analysis outputs, together with the repository revision at which they were written.

\section{Supplementary results}
\label{app:supp}

\subsection{Cutoff basins and population estimates}
\label{app:basins}
At most heights the cutoff is well determined: one basin takes most of the replicates, and the rule run on the full population selects the cutoff the sample does. Table~\ref{tab:basins} lists, for each height, the cutoff basins most often selected across the bootstrap replicates and the estimate obtained by running the same cutoff rule on every positive value of the population. One basin dominates at heights 300,000, 400,000 and 900,000 (83 to 99\% of replicates); at 500,000 three basins split the replicates, and at 200,000 the 170~BTC basin takes 76\% against 22\% for the 50~BTC atom. The population estimates agree with the sample fits within 0.6\% of the cutoff at six heights; at 900,000 the population prefers the 1~BTC basin (exponent 1.75) that the sample ranks second, and at 200,000 it moves further into the tail (299~BTC, exponent 2.13) than the dense sample (170~BTC, 2.07).

\begin{table}[htbp]
\centering
\caption{Cutoff basins and the population estimate. For each height the three basins most often selected across the 2,500 transaction-cluster replicates, with the fit on the observed sample (tail count, exponent, KS distance) and the share of replicates selecting each; $\ast$ marks the basin selected on the observed sample. The population row is the cutoff selected by the same rule on every positive value of the population, with its exponent and tail (Section~\ref{sec:curved}).}
\label{tab:basins}
\footnotesize\setlength{\tabcolsep}{5pt}
\begin{tabular}{@{}rlrrrr@{}}
\toprule
Height & Cutoff (BTC) & Tail $n$ & $\hat\alpha$ & KS & Selected (\%) \\
\midrule
200,000 & 169.8$\ast$ & 2,743 & 2.069 & 0.0755 & 76.2 \\
 & 50.0 & 7,869 & 1.945 & 0.0809 & 22.4 \\
 & 150.0 & 2,993 & 2.026 & 0.0948 & 1.3 \\
 & \emph{population} 298.9 & 3,638 & 2.127 & 0.0633 & \\
\addlinespace
300,000 & 0.0100$\ast$ & 313,071 & 1.463 & 0.0483 & 97.2 \\
 & 0.0122 & 271,632 & 1.437 & 0.0553 & 2.0 \\
 & 0.0492 & 162,236 & 1.495 & 0.0542 & 0.4 \\
 & \emph{population} 0.0100 & 3,233,721 & 1.453 & 0.0529 & \\
\addlinespace
400,000 & 1.021$\ast$ & 17,054 & 1.622 & 0.0439 & 99.2 \\
 & 0.0768 & 69,233 & 1.531 & 0.0498 & 0.8 \\
 & \emph{population} 1.000 & 620,342 & 1.655 & 0.0388 & \\
\addlinespace
500,000 & 0.0073 & 92,743 & 1.534 & 0.0372 & 27.8 \\
 & 0.0779 & 28,546 & 1.613 & 0.0375 & 27.5 \\
 & 1.002$\ast$ & 5,546 & 1.675 & 0.0364 & 27.0 \\
 & \emph{population} 1.000 & 719,281 & 1.681 & 0.0358 & \\
\addlinespace
600,000 & 1.010$\ast$ & 6,202 & 1.706 & 0.0355 & 73.0 \\
 & 0.0784 & 33,725 & 1.637 & 0.0404 & 11.8 \\
 & 0.0079 & 105,053 & 1.541 & 0.0398 & 9.6 \\
 & \emph{population} 1.002 & 812,344 & 1.725 & 0.0310 & \\
\addlinespace
700,000 & 0.0764 & 35,447 & 1.640 & 0.0361 & 53.6 \\
 & 1.018$\ast$ & 6,242 & 1.682 & 0.0349 & 46.3 \\
 & 1.155 & 5,671 & 1.674 & 0.0427 & 0.1 \\
 & \emph{population} 1.002 & 814,558 & 1.717 & 0.0340 & \\
\addlinespace
800,000 & 0.0819 & 46,121 & 1.688 & 0.0341 & 55.1 \\
 & 1.007$\ast$ & 7,633 & 1.761 & 0.0338 & 35.6 \\
 & 0.0800 & 46,690 & 1.685 & 0.0366 & 8.2 \\
 & \emph{population} 1.006 & 962,200 & 1.768 & 0.0318 & \\
\addlinespace
900,000 & 0.0825$\ast$ & 45,780 & 1.695 & 0.0310 & 83.3 \\
 & 1.006 & 7,389 & 1.740 & 0.0326 & 8.6 \\
 & 0.0799 & 46,590 & 1.692 & 0.0341 & 7.8 \\
 & \emph{population} 1.010 & 928,211 & 1.751 & 0.0305 & \\
\bottomrule
\end{tabular}
\end{table}

\subsection{Sensitivity to the floor and the basin count}
\label{app:sensitivity}
Neither of the procedure's two fixed constants carries the conclusion. The cutoff procedure fixes them by choice, at least 2,500 values above a candidate cutoff and 16 retained basins, and amendment v10 reruns basin identification and the 2,500-replicate bootstrap with one constant changed at a time, at heights 200,000 (dense sample), 500,000 and 900,000 (Table~\ref{tab:sensitivity}). The verdict holds in all twelve runs. Retaining 8 or 32 basins reproduces every reported number at every height, the winning basins being among the eight best; the KS ratio moves by at most 0.5, because the synthetic samples reselect among a different set. The floor changes nothing at 500,000 and 900,000, whose tails lie far above it.

At 200,000 the dense tail of 2,743 values sits close to the floor, and there the floor moves the cutoff the way the Methods anticipate. At a floor of 1,000 the rule retreats to 294.8~BTC (tail 1,588, exponent 2.109), the basin the population selects (Table~\ref{tab:basins}); at 5,000 it falls back to the 50~BTC atom (tail 7,869, exponent 1.945), the runner-up basin of the primary run. The exponent stays above 1.9 and the rejection is stronger under both. The outputs are in the data deposit from its version 1.1 and are mirrored in the repository.

\begin{table}[htbp]
\centering
\caption{Sensitivity of the cutoff procedure to its two fixed constants (amendment v10). The primary row is the run the paper reports; each variant changes one constant, the minimum number of values above a candidate cutoff or the number of KS-profile basins retained for reselection, and reruns basin identification and the 2,500-replicate transaction-cluster bootstrap with the primary seed. Interval: 95\% percentile interval of the exponent over replicates. KS ratio: observed KS distance over the median synthetic one. Share: percentage of replicates selecting the basin selected on the observed sample. Height 200,000 uses the dense v6 sample.}
\label{tab:sensitivity}
\footnotesize\setlength{\tabcolsep}{3.5pt}
\begin{tabular}{@{}lrrrrrlrrlr@{}}
\toprule
Height, variant & \shortstack[r]{Min.\\tail} & Basins & \shortstack[r]{Cutoff\\(BTC)} & \shortstack[r]{Tail\\count} & $\hat\alpha$ & 95\% interval & KS & \shortstack[r]{KS\\ratio} & $p$ & \shortstack[r]{Share\\(\%)} \\
\midrule
200,000, primary & 2,500 & 16 & 169.8 & 2,743 & 2.069 & [1.933, 2.106] & 0.0755 & 6.3 & 1/2501 & 76 \\
\quad floor-1000 & 1,000 & 16 & 294.8 & 1,588 & 2.109 & [2.023, 2.161] & 0.0624 & 3.9 & 1/2501 & 79 \\
\quad floor-5000 & 5,000 & 16 & 50.0 & 7,869 & 1.945 & [1.923, 1.967] & 0.0809 & 10.7 & 1/2501 & 99 \\
\quad basins-8 & 2,500 & 8 & 169.8 & 2,743 & 2.069 & [1.933, 2.106] & 0.0755 & 6.3 & 1/2501 & 76 \\
\quad basins-32 & 2,500 & 32 & 169.8 & 2,743 & 2.069 & [1.933, 2.106] & 0.0755 & 6.3 & 1/2501 & 76 \\
\addlinespace
500,000, primary & 2,500 & 16 & 1.002 & 5,546 & 1.675 & [1.526, 1.688] & 0.0364 & 11.1 & 1/2501 & 27 \\
\quad floor-1000 & 1,000 & 16 & 1.002 & 5,546 & 1.675 & [1.526, 1.688] & 0.0364 & 11.2 & 1/2501 & 27 \\
\quad floor-5000 & 5,000 & 16 & 1.002 & 5,546 & 1.675 & [1.526, 1.688] & 0.0364 & 11.2 & 1/2501 & 27 \\
\quad basins-8 & 2,500 & 8 & 1.002 & 5,546 & 1.675 & [1.526, 1.688] & 0.0364 & 10.7 & 1/2501 & 27 \\
\quad basins-32 & 2,500 & 32 & 1.002 & 5,546 & 1.675 & [1.526, 1.688] & 0.0364 & 11.2 & 1/2501 & 27 \\
\addlinespace
900,000, primary & 2,500 & 16 & 0.0825 & 45,780 & 1.695 & [1.687, 1.748] & 0.0310 & 10.2 & 1/2501 & 83 \\
\quad floor-1000 & 1,000 & 16 & 0.0825 & 45,780 & 1.695 & [1.687, 1.748] & 0.0310 & 10.2 & 1/2501 & 83 \\
\quad floor-5000 & 5,000 & 16 & 0.0825 & 45,780 & 1.695 & [1.687, 1.748] & 0.0310 & 10.1 & 1/2501 & 83 \\
\quad basins-8 & 2,500 & 8 & 0.0825 & 45,780 & 1.695 & [1.687, 1.748] & 0.0310 & 9.7 & 1/2501 & 83 \\
\quad basins-32 & 2,500 & 32 & 0.0825 & 45,780 & 1.695 & [1.687, 1.748] & 0.0310 & 10.4 & 1/2501 & 83 \\
\addlinespace
\bottomrule
\end{tabular}
\end{table}

\subsection{Value held by output size class}
\label{app:valuebands}

Table~\ref{tab:valuebands} gives, for each height, the share of the live coin supply held in outputs of each size class, computed exactly over every positive output; Figure~\ref{fig:valuebands} plots it together with the share of outputs.

\begin{table}[htbp]
\centering
\caption{Share of the live coin supply (\%) held in outputs of each size class, in BTC, at each height. Exact counts over every positive output.}
\label{tab:valuebands}
\footnotesize\setlength{\tabcolsep}{4pt}
\begin{tabular}{@{}rrrrrrrrrr@{}}
\toprule
Height & $<0.001$ & 0.001--0.01 & 0.01--0.1 & 0.1--1 & 1--10 & 10--100 & 100--1{,}000 & 1{,}000--10{,}000 & $\ge 10{,}000$ \\
\midrule
200,000 & 0.0 & 0.0 & 0.1 & 1.1 & 5.4 & 34.8 & 23.3 & 23.7 & 11.5 \\
300,000 & 0.0 & 0.0 & 0.5 & 1.9 & 6.7 & 31.3 & 25.9 & 22.7 & 11.0 \\
400,000 & 0.0 & 0.1 & 0.7 & 3.0 & 9.4 & 29.8 & 26.0 & 21.1 & 9.9 \\
500,000 & 0.0 & 0.2 & 1.3 & 4.4 & 10.3 & 28.1 & 26.3 & 21.4 & 7.9 \\
600,000 & 0.0 & 0.3 & 1.4 & 4.9 & 10.8 & 25.9 & 27.3 & 20.9 & 8.6 \\
700,000 & 0.0 & 0.3 & 1.4 & 4.8 & 10.4 & 25.4 & 28.3 & 21.1 & 8.2 \\
800,000 & 0.0 & 0.4 & 1.9 & 6.4 & 12.1 & 25.3 & 25.3 & 19.5 & 9.0 \\
900,000 & 0.1 & 0.3 & 1.8 & 6.3 & 11.5 & 24.6 & 28.8 & 18.4 & 8.3 \\
\bottomrule
\end{tabular}
\end{table}

\subsection{The alternatives fitted with cutoffs of their own}
\label{app:v7}
Neither the lognormal nor the cutoff power law survives the test when it is given a cutoff of its own. The comparison of Section~\ref{sec:results} fits every alternative on the tail selected for the power law, which is what a likelihood ratio requires. Amendment v7 asks the complementary question: treated as the primary candidate, with a cutoff selected by minimising its own KS distance over the same candidate grid and basins, and tested with its own semiparametric bootstrap, does either family pass? Table~\ref{tab:v7} gives the answer at heights 500,000 and 900,000. Both are rejected with $p = 1/2501$ and KS ratios between 6.8 and 19.6, of the same order as the pure power law's.

The cutoffs they choose are revealing. The lognormal's own cutoff lies far below the power law's, at 0.011~BTC at 500,000 and at 3,969 satoshis at 900,000, where its tail holds 43\% of all positive outputs: the lognormal is the best description of the bulk as well as of the tail, and still fails at these sample sizes. The cutoff power law selects 0.08~BTC at both heights. At 900,000 this is the power law's own basin, with the same tail; at 500,000 it is one of the power law's lower basins, twelve times below the 1~BTC cutoff the power law selects, with a tail five times larger. Its exponential scale of 700 to 1,300~BTC agrees with the fits on the power law's cutoff.

Both families tested this way fail on their own terms at both heights; the alternatives are better than the power law, not adequate. This is a statement about the two families tested at two heights under this procedure, not about every two-parameter family: the stretched exponential was not given its own cutoff, and families outside the set were not tried.

\begin{table}[htbp]
\centering
\caption{Each family fitted as the primary candidate, with its own KS-selected cutoff and its own semiparametric bootstrap of 2,500 transaction-cluster replicates (amendment v7), beside the pure power law of Table~\ref{tab:gof}. The cutoff power law (cutoff PL) is reported with its scale $x_{\min}/q$ in BTC, the value at which the exponential factor reaches $e^{-1}$. KS ratio is the observed KS distance over the median synthetic one.}
\label{tab:v7}
\footnotesize\setlength{\tabcolsep}{2.5pt}
\begin{tabular}{@{}rlrrrlrrr@{}}
\toprule
Height & Family & Cutoff (BTC) & Tail $n$ & Tail share & Parameters & KS & KS ratio & $p$ \\
\midrule
500,000 & power law & 1.0018 & 5,546 & 1.2\% & $\alpha=1.675$ & 0.0364 & 11.1 & 1/2501 \\
 & lognormal & 0.0110 & 71,200 & 15.2\% & $\mu=8.86$, $\sigma=4.07$ & 0.0147 & 6.8 & 1/2501 \\
 & cutoff PL & 0.0823 & 27,815 & 5.9\% & $\alpha=1.593$, scale 683 & 0.0292 & 7.8 & 1/2501 \\
\addlinespace
900,000 & power law & 0.0825 & 45,780 & 3.5\% & $\alpha=1.695$ & 0.0310 & 10.2 & 1/2501 \\
 & lognormal & 0.000040 & 566,952 & 42.9\% & $\mu=9.78$, $\sigma=3.76$ & 0.0162 & 19.6 & 1/2501 \\
 & cutoff PL & 0.0829 & 45,697 & 3.5\% & $\alpha=1.685$, scale 1,347 & 0.0286 & 9.6 & 1/2501 \\
\bottomrule
\end{tabular}
\end{table}

\subsection{Direct comparison between the curved alternatives}
\label{app:paired}
The lognormal and the cutoff power law differ clearly at only two heights. Table~\ref{tab:paired} compares the alternatives with each other on the selected tail, replicate by replicate, as the per-observation difference of their log-likelihoods with a 95\% percentile interval over the transaction-cluster replicates. At six heights the interval covers zero and no median exceeds 0.0013 nats per observation, so the data give no clear evidence of a difference between the two there. At 300,000 and 400,000 they are separated: the cutoff power law is ahead by 0.0027 and 0.0033 nats with intervals that exclude zero. That is a difference at those two heights, not an ordering of the two families over the panel, and the held-out comparison of Table~\ref{tab:alternatives} agrees with the in-sample ranking at all eight. The stretched exponential is ahead of the lognormal at every height, by at most 0.0014 nats, with intervals that exclude zero from 200,000 to 600,000.

\begin{table}[t]
\centering
\caption{Direct comparison between the curved alternatives on the selected tail: log-likelihood ratio per tail observation, first model minus second (positive favours the first), median and 95\% percentile interval over transaction-cluster replicates.}
\label{tab:paired}
\footnotesize\setlength{\tabcolsep}{4pt}
\begin{tabular}{rlll}
\toprule
Height & Lognormal $-$ cutoff PL & Lognormal $-$ stretched exp. & Stretched exp. $-$ cutoff PL \\
\midrule
200,000 & -0.0004 [-0.0013, +0.0006] & -0.0001 [-0.0002, -0.0000] & -0.0003 [-0.0012, +0.0007] \\
300,000 & -0.0027 [-0.0037, -0.0012] & -0.0003 [-0.0005, -0.0001] & -0.0024 [-0.0034, -0.0008] \\
400,000 & -0.0033 [-0.0058, -0.0008] & -0.0014 [-0.0017, -0.0011] & -0.0019 [-0.0042, +0.0005] \\
500,000 & +0.0004 [-0.0055, +0.0014] & -0.0002 [-0.0007, -0.0000] & +0.0005 [-0.0049, +0.0015] \\
600,000 & -0.0029 [-0.0048, +0.0013] & -0.0003 [-0.0005, -0.0000] & -0.0027 [-0.0045, +0.0015] \\
700,000 & +0.0013 [-0.0028, +0.0024] & -0.0000 [-0.0004, +0.0000] & +0.0014 [-0.0025, +0.0026] \\
800,000 & +0.0008 [-0.0034, +0.0017] & -0.0000 [-0.0002, +0.0000] & +0.0008 [-0.0032, +0.0017] \\
900,000 & +0.0006 [-0.0022, +0.0012] & -0.0000 [-0.0001, +0.0000] & +0.0007 [-0.0021, +0.0012] \\
\bottomrule
\end{tabular}
\end{table}

\subsection{Held-out predictive comparison}
\label{app:heldout}
Against the pure power law every alternative predicts held-out transactions better at every height but one, in aggregate and in four or five folds out of five (Table~\ref{tab:heldout}). Against each other the lognormal and the cutoff power law split four heights each, the same way the in-sample comparison ranks them. The exception is the cutoff power law at 700,000, which is ahead in sample by 0.0007 nats and behind out of sample by 0.0027: the height where its fitted scale is set by a single very large output is also the height where it generalises worst.

\begin{table}[htbp]
\centering
\caption{Five-fold held-out predictive comparison, conditional on the basin set identified on the whole sample. Transactions are split into five folds by an independent byte of their identifier; each model is fitted on four folds at the cutoff selected there and scored on the fifth. Entries are the held-out log score minus the pure power law's, in nats per held-out observation, with the number of folds favouring the alternative in parentheses; the last column repeats the lognormal against the cutoff power law. A larger family pays out of sample for the parameter it gains in sample, so a positive entry is evidence that the shapes differ. It is not proof: a positive score can arise by chance, and the candidate cutoffs were identified on the whole sample.}
\label{tab:heldout}
\footnotesize\setlength{\tabcolsep}{5pt}
\begin{tabular}{@{}rrllll@{}}
\toprule
 & & \multicolumn{3}{c}{Held out, alternative minus pure power law} & \\
\cmidrule(lr){3-5}
Height & Held out $n$ & Lognormal & Stretched exp. & Cutoff PL & Lognormal $-$ cutoff PL \\
\midrule
200,000 & 7,869 & +0.0014 (4/5) & +0.0015 (4/5) & +0.0017 (4/5) & -0.0003 \\
300,000 & 313,071 & +0.0017 (4/5) & +0.0020 (4/5) & +0.0045 (5/5) & -0.0028 \\
400,000 & 17,054 & +0.0099 (5/5) & +0.0113 (5/5) & +0.0129 (5/5) & -0.0030 \\
500,000 & 66,834 & +0.0036 (5/5) & +0.0038 (5/5) & +0.0031 (5/5) & +0.0006 \\
600,000 & 6,202 & +0.0020 (5/5) & +0.0023 (5/5) & +0.0054 (5/5) & -0.0033 \\
700,000 & 23,783 & +0.0025 (5/5) & +0.0026 (5/5) & -0.0027 (4/5) & +0.0053 \\
800,000 & 38,493 & +0.0029 (5/5) & +0.0029 (5/5) & +0.0019 (5/5) & +0.0010 \\
900,000 & 45,780 & +0.0017 (5/5) & +0.0017 (5/5) & +0.0009 (4/5) & +0.0007 \\
\bottomrule
\end{tabular}
\end{table}

\subsection{Supply extrapolation of the alternatives}
\label{app:supplyalt}
The fitted alternatives stay far closer to the supply than the pure power law. Table~\ref{tab:supplyalt} repeats the extrapolation of Table~\ref{tab:supply} for them, over the same population tail. The lognormal and the stretched exponential still over-extrapolate the number of outputs above the largest existing one, by one to two orders of magnitude, but they place fewer than one expected output above the supply held in the UTXO set at seven of the eight heights, and 14 and 5 respectively at 300,000; the cutoff power law places far less than one output above the supply at every height and at most one above the largest existing output. Expected counts below one are not zero, and the very small numbers in the table are the value of an extrapolation, not a claim that the fitted family has bounded support. The failure to respect the supply is specific to the pure law.

\begin{table}[t]
\centering
\caption{Extrapolation of each fitted tail model over the population tail: expected number of outputs above the largest existing output / above the supply mined at the height. Fits from the observed samples above the selected cutoff; population tail as in Table~\ref{tab:supply}.}
\label{tab:supplyalt}
\footnotesize\setlength{\tabcolsep}{4pt}
\begin{tabular}{rllll}
\toprule
Height & Pure power law & Lognormal & Stretched exponential & Cutoff power law \\
\midrule
200,000 & 9 / 0.05 & 4 / 2.5e-03 & 4 / 1.3e-03 & 1 / 3.7e-74 \\
300,000 & 2,127 / 204 & 514 / 14 & 344 / 5 & 4.5e-37 / 0.0e+00 \\
400,000 & 485 / 21 & 14 / 4.9e-03 & 2 / 1.6e-06 & 1.3e-46 / 0.0e+00 \\
500,000 & 345 / 9 & 34 / 0.03 & 14 / 1.1e-03 & 1.5e-35 / 0.0e+00 \\
600,000 & 213 / 6 & 30 / 0.08 & 16 / 0.01 & 7.1e-39 / 0.0e+00 \\
700,000 & 229 / 9 & 31 / 0.13 & 19 / 0.03 & 0.06 / 3.9e-304 \\
800,000 & 99 / 3 & 16 / 0.06 & 10 / 0.02 & 7.0e-63 / 0.0e+00 \\
900,000 & 288 / 9 & 20 / 0.05 & 13 / 0.01 & 7.4e-43 / 0.0e+00 \\
\bottomrule
\end{tabular}
\end{table}

\subsection{Calibration of the procedure under dependence}
\label{app:calibration}

The semiparametric null draws the synthetic tail values independently given the resampled transactions, so dependence among the tail values of one transaction is present in the data and absent from the null. Two things are needed to judge what that costs: how much such dependence the fitted tails contain, and what it does to the level of the test.

Table~\ref{tab:dependence} measures the first. The share of tail outputs that sit in a transaction with another tail output ranges from 3\% at the height-200,000 stock to 58\% at 300,000 and 72\% in the flow window at 700,000, and the correlation of their normal scores, the quantity the simulation below varies, is 0.50 to 0.90 on the stock tails and 0.10 to 0.54 on the flow tails. The estimator returns 0.02, 0.62, 0.89 and 1.00 on synthetic samples built at 0, 0.5, 0.9 and 1, so the real tails sit between the second and third rows of the calibration and nowhere near the fourth.

\begin{table}[htbp]
\centering
\caption{Within-transaction dependence of the tails actually fitted: the share of tail outputs in transactions holding two or more of them, the largest such transaction, and the correlation $\rho$ of their van der Waerden scores, which is the quantity the calibration of Section~\ref{app:calibration} varies. On synthetic samples built at $\rho = 0$, 0.5, 0.9 and 1 the same estimator returns 0.02, 0.62, 0.89 and 1.00.}
\label{tab:dependence}
\footnotesize\setlength{\tabcolsep}{4.5pt}
\begin{tabular}{@{}rrrrr rrrr@{}}
\toprule
 & \multicolumn{4}{c}{Stock} & \multicolumn{4}{c}{Flow} \\
\cmidrule(lr){2-5}\cmidrule(lr){6-9}
Height & Tail $n$ & In txs $\ge 2$ (\%) & Largest tx & $\rho$ & Tail $n$ & In txs $\ge 2$ (\%) & Largest tx & $\rho$ \\
\midrule
200,000 & 2,743 & 3.1 & 20 & 0.65 & 78,861 & 14.7 & 25 & 0.31 \\
300,000 & 313,071 & 57.7 & 1,985 & 0.57 & 3,653 & 6.5 & 10 & 0.44 \\
400,000 & 17,054 & 10.7 & 42 & 0.75 & 126,100 & 37.5 & 126 & 0.37 \\
500,000 & 5,546 & 10.9 & 20 & 0.63 & 13,801 & 8.8 & 26 & 0.50 \\
600,000 & 6,202 & 10.1 & 25 & 0.90 & 3,231 & 11.6 & 21 & 0.54 \\
700,000 & 6,242 & 8.0 & 20 & 0.56 & 178,233 & 71.9 & 4,188 & 0.10 \\
800,000 & 7,633 & 8.2 & 10 & 0.50 & 167,305 & 67.1 & 1,775 & 0.11 \\
900,000 & 45,780 & 18.2 & 192 & 0.56 & 251,444 & 65.2 & 1,735 & 0.46 \\
\bottomrule
\end{tabular}
\end{table}

Amendment v8 measures the second. Synthetic samples keep the transaction structure and the body values of a height-200,000 sample and replace every tail value by a draw from the fitted discrete power law through a Gaussian copula with within-transaction correlation $\rho \in \{0, 0.5, 0.9, 1\}$, where $\rho = 0$ reproduces the null construction and $\rho = 1$ makes every tail output of a transaction identical. One hundred samples per $\rho$ are analysed with the frozen scripts and 1,000 replicates each; a secondary run repeats $\rho = 0.9$ on the height-500,000 structure. Samples on which a cluster resample left fewer than 2,500 tail observations in every retained basin, so that the procedure stopped, are reported as aborted and counted neither as rejections nor as non-rejections.

Two limits of this design should be read with its results. The structure is the 0.78\% transaction sample of height 200,000, which Section~\ref{sec:robustness} supersedes at that height: it is not the million-output sample the panel now uses at 200,000 to 400,000. From 500,000 onward the 0.78\% sample and the dense design are the same sample, verified output by output, so the secondary run is on the structure the panel actually uses at that height, but it has only ten data sets. And a maximum over a finite set of simulations is not a bound: these figures describe the Gaussian-copula construction at the correlations simulated, not every possible form of dependence.

Table~\ref{tab:calibration200k} gives the result on the height-200,000 structure, with exact binomial intervals. With independent tail values the procedure rejects 6\% of true power-law samples at the nominal 10\% level (95\% interval 2 to 13\%) and 3\% at 5\%: the level is held, and if anything the test is slightly conservative, as expected when duplicated transactions widen the synthetic reference. Moderate dependence leaves the level at 5\% (1 to 11\%). Strong dependence inflates it, to 16\% (9 to 25\%) at $\rho = 0.9$ and 57\% (46 to 67\%) when every tail output of a transaction is identical. The exponent is recovered without bias at every $\rho$ (mean 1.530 to 1.534 against a true value of 1.532).

The KS ratio stays below 2.3 under all correlations up to 0.9 and below 2.9 at $\rho = 1$, against observed ratios of 5 to 42 on the stock and 2.7 to 23 on the flow; the two weakest flow rejections, at 600,000 and 300,000, are the only ones inside that range, and both were therefore repeated against the dependence-preserving null (Section~\ref{sec:robustness}). Between 1 and 12 samples per $\rho$ aborted at the eligibility boundary: on those every retained basin had a tail within a few hundred observations of the 2,500 floor, so a cluster resample that lost a few tail outputs left no eligible basin. The height-200,000 tail of the current analysis holds 2,743 values against the same floor, which is closer to it than any other height in the panel.

The secondary run on the height-500,000 structure (Table~\ref{tab:calibration500k}), with 216,000 transactions and a tail near 1~BTC, rejects none of ten samples at $\rho = 0.9$, with a KS ratio never above 1.2 and the exponent recovered at 1.672 against a true 1.675. Ten samples put a 95\% upper bound of 31\% on the rejection probability, so this shows the inflation seen on the small early structure does not reappear here; it does not by itself establish a 10\% level at that height.

Because neither run certifies the panel, Section~\ref{sec:robustness} repeats the test at every height against a null built to carry the measured dependence. That is the more direct instrument, and it agrees with these simulations about where dependence matters: the height whose KS ratio it moves is the height with the most dependence in its tail.

\begin{table}[htbp]
\centering
\caption{Gains of the curved families on tails that really are power laws. Each synthetic sample of Appendix~\ref{app:calibration} has a true power-law tail on a real transaction structure; the three families are fitted on it at its own selected cutoff, exactly as on the data. Entries are the 95th percentile and the maximum of the improvement in total log-likelihood over the power law, in nats, across the data sets of that row. These are ranges observed in particular simulations, not thresholds with a known error rate: exceeding them is informative, and falling inside them is not evidence that a tail is a power law.}
\label{tab:benchmark}
\footnotesize\setlength{\tabcolsep}{4pt}
\begin{tabular}{@{}llrr rr rr rr@{}}
\toprule
 & & & & \multicolumn{2}{c}{Lognormal} & \multicolumn{2}{c}{Stretched exp.} & \multicolumn{2}{c}{Cutoff PL} \\
\cmidrule(lr){5-6}\cmidrule(lr){7-8}\cmidrule(lr){9-10}
Structure & $\rho$ & Sets & Median tail & 95\% & Max & 95\% & Max & 95\% & Max \\
\midrule
200,000 (0.78\% sample) & 0.00 & 99 & 3,405 & 1.5 & 3.8 & 1.5 & 3.6 & 1.8 & 4.7 \\
 & 0.50 & 88 & 3,410 & 1.5 & 2.6 & 1.5 & 2.7 & 2.1 & 3.5 \\
 & 0.90 & 95 & 3,417 & 1.9 & 4.8 & 1.9 & 5.1 & 2.0 & 5.2 \\
 & 1.00 & 93 & 3,416 & 2.8 & 4.2 & 2.8 & 4.3 & 2.8 & 6.2 \\
500,000 & 0.90 & 10 & 5,514 & 0.7 & 1.2 & 0.8 & 1.3 & 1.0 & 1.0 \\
\bottomrule
\end{tabular}
\end{table}

\IfFileExists{sections/table_calibration_200k.tex}{
\begin{table}[t]
\centering
\caption{Calibration of the goodness-of-fit procedure on synthetic samples with a true power-law tail and within-transaction dependence $\rho$ (Gaussian copula) on the height-200k cluster structure. Rejection frequencies among completed data sets at nominal levels 0.10 and 0.05, median $p$, KS ratio (observed over median synthetic) quantiles, mean fitted exponent. ``Aborted'' counts data sets on which the procedure stopped because a cluster resample left fewer than 2,500 tail observations in every retained basin.}
\label{tab:calibration200k}
\footnotesize\setlength{\tabcolsep}{3pt}
\begin{tabular}{@{}rrrrrrrrrr@{}}
\toprule
$\rho$ & Completed & Aborted & Reject 0.10 & Reject 0.05 & Median $p$ & \multicolumn{3}{c}{KS ratio: median, 95\%, max} & Mean $\hat\alpha$ \\
\midrule
0.0 & 99 & 1 & 0.06 & 0.03 & 0.61 & 0.92 & 1.43 & 1.88 & 1.534 \\
0.5 & 88 & 12 & 0.05 & 0.05 & 0.61 & 0.92 & 1.38 & 1.95 & 1.530 \\
0.9 & 95 & 5 & 0.16 & 0.14 & 0.47 & 1.02 & 1.87 & 2.29 & 1.530 \\
1.0 & 93 & 7 & 0.57 & 0.43 & 0.08 & 1.45 & 2.38 & 2.85 & 1.532 \\
\bottomrule
\end{tabular}
\end{table}
}{}
\IfFileExists{sections/table_calibration_500k.tex}{
\begin{table}[t]
\centering
\caption{Calibration of the goodness-of-fit procedure on synthetic samples with a true power-law tail and within-transaction dependence $\rho$ (Gaussian copula) on the height-500k cluster structure. Rejection frequencies among completed data sets at nominal levels 0.10 and 0.05, median $p$, KS ratio (observed over median synthetic) quantiles, mean fitted exponent. ``Aborted'' counts data sets on which the procedure stopped because a cluster resample left fewer than 2,500 tail observations in every retained basin.}
\label{tab:calibration500k}
\footnotesize\setlength{\tabcolsep}{3pt}
\begin{tabular}{@{}rrrrrrrrrr@{}}
\toprule
$\rho$ & Completed & Aborted & Reject 0.10 & Reject 0.05 & Median $p$ & \multicolumn{3}{c}{KS ratio: median, 95\%, max} & Mean $\hat\alpha$ \\
\midrule
0.9 & 10 & 0 & 0.00 & 0.00 & 0.67 & 0.89 & 1.10 & 1.19 & 1.672 \\
\bottomrule
\end{tabular}
\end{table}
}{}

\FloatBarrier
\section{AI use disclosure}
\label{app:ai}

The analysis and the manuscript were produced with AI assistance. Claude Fable~5.1 and Claude Opus~5
(Anthropic), used interactively through Claude Code, implemented the UTXO replay, the fitting,
bootstrap and calibration code, the figures and tables, the verification tools described below, and
drafts of the text. The authors set the question, the estimand and the protocol, and made every
scientific decision: the panel of heights, the cutoff rule and its amendments, the treatment of
dependence, the decision rules, and the interpretation of the results. The protocol was fixed before
any inference and every later change is a dated amendment (Appendix~\ref{app:protocol}).

One of the checks is itself model-based and is deposited with the code: every numeric sentence in the
manuscript was matched against the recorded analysis outputs, which caught two stale figures that
earlier rounds of review had missed.

The manuscript, the protocol documents and the analysis code were also reviewed in six rounds by
\texttt{gpt-6-astra} (OpenAI), an AI system independent of the one that supported the experiment, which modified nothing. The
reviews and the responses to them are in the repository; they led among other corrections to the
lognormal parameterisation of amendment~v9 and to the removal of an unsupported attribution.

The authors take full responsibility for the content and for the correctness of the results. The AI
tools were assistants and do not meet the criteria for authorship; every output they produced was
checked against the recorded analysis or the cited sources before use.

\end{document}